\documentstyle[epsf,psfig]{mn}

\newcommand{\SS}{\scriptscriptstyle}
\newcommand{\Rl}{R_{\SS L1}}
\newcommand{\Md}{\dot{\cal M}}
\newcommand{\SC}{\scriptsize}
\newcommand{\Teff}{T_{\mbox{\SC eff}}}
\newcommand{\Myr}{{\cal M}_\odot\!\mbox{yr}^{-1}}

\newcommand{\gdot}{^\circ_{\raisebox{.6ex}{\hspace{.05em}.}}}
\newcommand{\gcm}{g$\,$cm$^{-2}$}
\newcommand{\gsim}{{\textstyle{\; \lower 0.7ex\hbox{$>$}\; 
\atop \raise-0.1ex\hbox{$\sim$}}}}
\newcommand{\lsim}{{\textstyle{\; \lower 0.7ex\hbox{$<$}\; 
\atop \raise-0.1ex\hbox{$\sim$}}}}

\begin{document}

\title{The Patchy Accretion Disc in HT Cassiopeiae}

\author[Sonja~Vrielmann et al.]
{Sonja~Vrielmann$^1$\thanks{Send offprint requests to:
sonja@pinguin.ast.uct.ac.za},
Frederic~V.~Hessman$^2$,
Keith~Horne$^3$,\\
$^1$Department of Astronomy, University of Cape Town, Private Bag,
Rondebosch, 7700, South Africa\\
$^2$Universit\"ats-Sternwarte G\"ottingen, Geismarlandstr.\ 11, 37083
G\"ottingen, Germany\\
$^3$School of Physics and Astrophysics, North Haugh, St.\ Andrews,
Fife, KY16 9SS, Scotland\\
}

\maketitle

\begin{abstract}

We have reconstructed the temperatures and surface densities in the
quiescent accretion disc in HT~Cas by performing a {\em Physical
Parameter Eclipse Mapping} analysis of archival UBVR observations.
Using a simple hydrogen slab model and demanding a smooth, maximally
artefact-free reconstruction, we derive a formal distance to HT~Cas of
$207\pm 10$\,pc, significantly larger than the $133\pm 14$\,pc we
derive from a re-analysis of the data in the literature.

The accretion disc is small ($0.3-0.4~\Rl$) and moderately optically
thin but becomes nearly optically thick near the white dwarf.
The temperatures and surface densities in the
disc range from 9\,500\,K and 0.013\,g\,cm$^{-2}$ in the center to about
4\,000\,K and 0.04\,g\,cm$^{-2}$ at the disc edge.  The mass-accretion
rate in the disc is roughly constant but -- at the derived distance --
uncomfortably close to those which would prohibit the dwarf nova
eruptions.

We argue that the larger derived distance is probably incorrect but is
not produced by inaccuracies in our spectral model or optimization
method.  The discrepancy can be resolved if the emission regions on
the disc are patchy with a filling factor of about 40\% of the disc's
surface.  This solves the problem with the high effective temperatures
in the disc -- reducing them to around 6\,500\,K within a radius of
$0.2\,\Rl$ -- and reduces the derived temperature of the white dwarf
and/or boundary layer from 22\,600 to 15\,500\,K.

The viscosity parameters $\alpha$ derived from all reconstructed
temperatures and surface densities are of order 10-100
and cannot be lowered significantly by invoking a lower distance or
the filling factor.   This situation is easily explained using
the same patchy nature of the emitting material, since the quiescent disc
cannot consist of optically thin regions alone, but also of a dark and
hence cold and dense disc which could easily contain most of the matter.
If we require global values of $\alpha$ of order 0.1, the implied
total surface densities are 1-100\,\gcm\ -- just like those expected
for quiescent discs awaiting the next eruption.

We discuss several possible sources of the chromospheric emission
and its patchiness, including irradiation of the disc, thermal
instabilities, spiral-wave-like global structures,
and magnetically active regions associated with dynamo action
and/or Balbus-Hawley instabilities.

\end{abstract}



\section{Introduction}

Cataclysmic variables (CVs), are semi-detached binary stars consisting
of a Roche lobe-filling late-type dwarf (the secondary star) which
loses matter onto a white dwarf (the primary). Unless the primary has
a considerable magnetic field, the transfered matter has enough
specific angular momentum to create a gaseous accretion disc around
this accreting star. Some CVs -- the dwarf novae -- have discs
which occasionally have luminous phases of high mass-accretion but
spend most of their time in a faint quiescent state.  In quiescence,
the optical light is usually dominated by optically thin line emission
from the disc and continuum emission from the white dwarf, the disc
and the bright spot caused by the impact of the transferred material
onto the disc.

While spectral and eclipse observations of erupting dwarf novae show
that the disc is optically thick during the outburst state, there is
considerable uncertainty about the state of the quiescent discs.
Theoretical models for the outbursts invoking a thermal instability as
the cause of the eruption require that the quiescent disc store up
material for the next eruption in a less viscous state in which the
disc should be cool and optically thick (e.g. Ludwig et al. 1994).
The inner disc may be emptied by irradiation from the hot white dwarf
(Leach et al. 1999) or by a siphon flow fueled by a hot corona in the
inner disc (Meyer \& Meyer-Hofmeister 1994).  Thus, the expectation is
that the outer disc is cool and optically thick and the inner disc is
warm/hot and optically thin.

HT~Cas was once called ``the {\em Rosetta Stone} of dwarf novae'',
because -- being one of the relatively few eclipsing dwarf novae -- it
more readily revealed its secrets (Patterson 1981).  It has seldom been
caught in outburst because of the unusually long outburst period of
about 400 $\pm$ 50 days (Wenzel 1987). The quiescent periods may even
last up to almost 9 years\footnote{Since it is a circumpolar object
for many northern observatories, the likelihood of detecting outbursts
is relatively high.}  indicating an extremely low mass accretion rate
in quiescence.  The accretion disc is hardly visible during quiescence
and the bright spot, the impact region of the accretion stream from
the secondary, is nearly absent (Wood, Horne \& Vennes et al.\ 1992,
hereafter WHV92).  The light curves predominantly show the eclipse of
the white dwarf.  The optical spectrum of HT~Cas (Young, Schneider \&
Shectman 1981) shows double peaked emission lines of H~I, He~I, He~II
and Ca~II, typical for a high inclination system ($i = 81^\circ$:
Horne, Wood \& Stiening 1991, hereafter HWS91). Consistent with the
broad-band photometry, the averaged profiles do not show any
significant blue/red asymmetry, which would normally be attributed to
the bright spot.

WHV92 investigated the accretion disc using multicolour eclipse
photometry from which the contribution of the white dwarf had been
subtracted. They produced independent classical eclipse maps for each
colour and then modeled them with a two-parameter model spectrum
yielding the temperature and surface density distribution in the disc.
While this analysis suggested that the disc is optically thin -- as
expected -- it suffers in detail due to the individual smearing of
each eclipse map (see Vrielmann, Horne \& Hessman 1999; hereafter
VHH99). Furthermore, the pre-subtraction of the white dwarf contribution
and the use of simple Cartesian reconstruction maps can make it
difficult to produce a consistent physical picture of the various
components.

\begin{figure}
\psfig{file=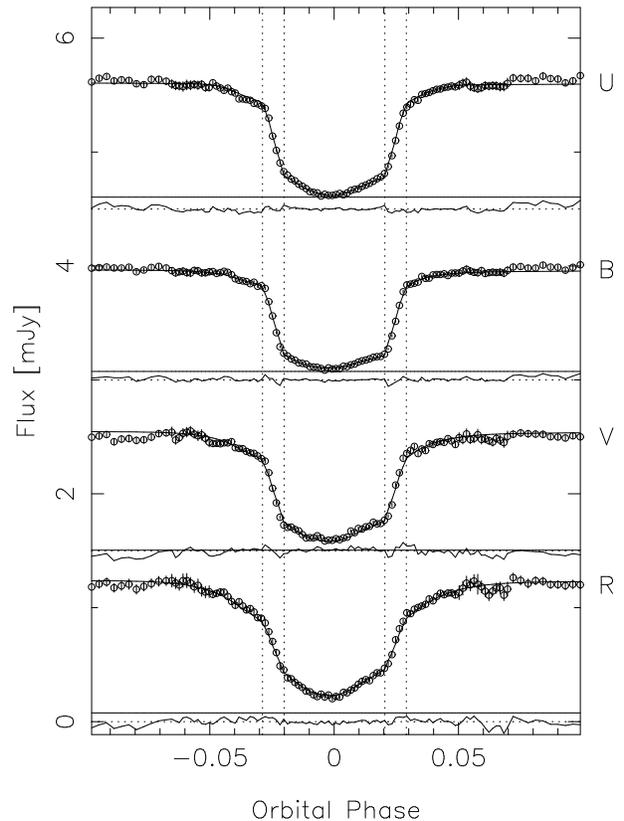,width=8cm}
\caption{\small The averaged quiescent light curves in UBVR together
with the fits made simultaneously in all four filters.
The contact phases of the white dwarf as given by Wood,
Horne \& Vennes (1992) are drawn as vertical dashed lines.
The light curves are offset by 1.5 mJy, the horizontal dotted lines
give the zero level for each light curve. In relation to this, the
uneclipsed components and the residuals are plotted (solid lines).
\label{htf}}
\end{figure}

Our present investigation is aimed at improving our understanding of
the physical state of the quiescent accretion disc in HT~Cas.  We
apply the {\it Physical Parameter Eclipse Mapping} method (PPEM) -- a
distinctly different tomographic ansatz (VHH99) which also uses the
Maximum-Entropy Method (MEM; Skilling \& Bryan 1984) -- to the same
UBVR photometry and derive the white dwarf temperature and the disc
structure by simultaneously fitting all available eclipse light
curves.


\section{The data and system parameters}

Since the dataset has been previously published and analysed, details
about the acquisition can be found in HWS91. The photometric data
consist of 16 eclipses taken in September 1982 and November/December 1983 at
the Palomar 60-inch telescope using the Stiening photometer which
gathers data in four filters simultaneously. The previous outburst
occured in March 1980 and the quiescence lasted until the super
outburst in January 1985. HT Cas was therefore in the middle
of a quiescence period.

The original data have a time resolution of 1~s. As can be seen in
Fig.\,1 in HWS91 the basic eclipse shape appears to be constant, we do
not even see a distiguished difference between the 1982 and 1983
data. But the individual eclipses suffer heavily from flickering. We
averaged all light curves in order to enhance any stable patterns and
minimize all random features, like flickering.
The averaged light curves were then
phase-binned to $\Delta \varphi = 0.0015$ near the eclipse (within
phases $-0.07$ and 0.07) and $\Delta \varphi = 0.003$ outside this
range (Fig.~\ref{htf}).

We used the photometric system parameters as given by HWS91, namely a
mass ratio $q = {\cal M}_{rd}/{\cal M}_{wd} = 0.15$, an inclination
angle $i = 81^\circ$, as well as a white dwarf mass and radius of
${\cal M}_{wd} = 0.61 {\cal M}_\odot$ and $R_{wd} = 0.0118 {\cal
R}_\odot = 0.0263\Rl$.  With an orbital period of 106\,min, the
distance $R_{L_1}$ from the primary to the L$_1$-point is $3.13\times
10^{10}$\,cm.  The mass and radius of the secondary star derived by
HWS91 are $0.09\,{\cal M}_\odot$ and $0.154\,{\cal R}_\odot$.

WHV92 determined the $E(B-V)$ as definately smaller than 0.2 and more
likely between 0 and 0.1. Therefore, we neglect the interstellar
reddening for our calculations.


\section{Distance estimates}
\label{distance}

Wood et al.\ (1995),
using UBVR photometry of the white dwarf extracted from the
eclipse light curve, derived a value of 165$\pm$10 pc.  Marsh (1990)
deduced a distance of 140$\pm$14 pc using the observed Na I and TiO
absorption spectrum of the secondary and his own calibration of the
TiO $\lambda\lambda 7165,7665$ band strengths, a surface brightness
calibration for M giants (the Barnes-Evans relation; Barnes et
al. 1978), and the radial velocity amplitude $K_2$.  This derivation
makes several assumptions about the nature of the secondary star.
Marsh predicted a mid-eclipse infrared H magnitude of $14.9\pm0.1$ for
the secondary but Berriman et al.\ (1987) measured $15.7\pm0.2$,
indicating that the system may be farther away than 200 pc.  Two
months after the first infrared measurements, Berriman et al.\
measured a much brighter H magnitude of 14.5, apparently confirming
Marsh's distance, but this difference could have been due to an
increase in the constribution from the uneclipsed component (e.g.\ the
outer regions of the disc).  Although Zhang et al.\ (1986) derived an
even larger distance of 320\,pc from their analysis of BVR light
curves, this distance is undoubtably made too large by their
assumption of pure blackbody emission from the disc and the lack of
U-band data.

Marsh's estimate depends on his identification of the M5.4 $\pm$ 0.03
spectrum (derived from his own calibration of the TiO band strengths),
the secondary's flux at mid-eclipse ($0.224 \pm 0.01$\,mJy at
7\,500\,{\AA}), and the Barnes-Evans relation (the surface brightness
of giant stars versus V-R colour); Barnes et al.\ (1978).  Beuermann
et al.\ (1999) have recalibrated the Barnes-Evans relations (surface
brightness versus colour) for both giants and dwarfs and find that the
relation for M5 dwarfs is about 0.12 mag fainter than that for giants.
This effect implies a 5\% decrease in Marsh's distance estimate
($132\pm13$ instead of $140\pm 14$\,pc).  Marsh's spectral type
depends upon the relative strengths of the TiO features at
$\lambda\lambda7165,7665$: the latter band is contaminated with weak
OI emission, so that the $\lambda7165/\lambda7665$ ratio may be
over-estimated, and the spectral type later than M5.4. Though, the
type is probably not any later than M6, since the region around TiO
$\lambda7165$ does not show the multiple absorbtion features prominent
for later types.

Weichhold \& Beuermann (2000) and Beuermann (2000) have also
constructed a TiO surface brightness relation which can be used for
systems like HT~Cas with severe veiling.  Using Marsh's M5.4 $\pm$ 0.3
spectral type and assuming $M_{rd} = 0.09\,M_\odot$, the observed TiO
flux at mid-eclipse and a 94\% correction for the Roche flattening at
phase 0.0 implies a distance of $135\pm14$\,pc, in agreement with our
revision of Marsh's estimate.  If the spectral type is as late as M6,
this number drops to 93\,pc, since the flux of the TiO bands drops
rapidly after M5.5.

One can reverse the calculation and compute the expected spectral
type given the observed TiO flux and assumed distance of 165~pc (Wood
et al. 1995): this results in a spectral type of M4.5.  This would imply an increase in the
$\lambda7165/\lambda7665$ ratio of 43\% over that measured by Marsh
(1990), a ratio which is already too large due to OI contamination.
A spectral type as early as M2-M3 (corresponding
to a distance of $\sim$200\,pc) can be simply ruled out because it would require twice as
much light at mid-eclipse as is available for the secondary and the
residual disc. The only way to change these results -- given that the
constraints come from the unilluminated backside of the secondary star
-- is to decrease the strength of the TiO band by lowering the
metalicity to at least $[M/H]\approx -1.0$ (to get a reduction of the
band depth by a factor $(133/200)^2=44\%$).  However, there is no
evidence to date of any low-metalicity CV's (see Beuermann et
al. 1998).  An earlier spectral type is also implausible due to the
fact that there is weak evidence that the spectral types of secondary
stars in CV's have {\it later} -- not {\it earlier} -- spectral types
than normal main-sequence stars (Beuermann et al.\ 1998).
        
We must conclude that Marsh's observations clearly suggest a distance
of $133\pm 14$\,pc (weighted average of $132\pm13$ and
$135\pm14$\,pc).  This result remains in conflict with the
eclipse-mapping results of Wood et al.\ (1995).  Given that the latter
depends upon several assumptions about the visibility of the white
dwarf and its temperature, and that the plausible corrections to our
re-analysis of Marsh's result tend to make the distance smaller, not
larger, we feel that our revision of Marsh's estimate is still more
reliable.  Still, given the uncertainties, it is possible that the
distance is somewhat larger.

In principle, the PPEM method can provide us with an independent
estimate of the distance, as previously shown by Vrielmann, Stiening
\& Offutt (2002) if a proper spectral model is used. The use of an
improper distance produces characteristic features in the maps of the
fitted quantities. In Appendix~\ref{app_dist} we demonstrate that
reliable distances can be found by maximizing the entropy of the fit
(making the disk maps as simple as possible) at fixed $\chi^2$ (requiring
a good fit to the lightcurves).  When the wrong distance is used, the
disk maps become more complicated, and the entropy accordingly becomes
increasingly negative.  The best distance is thus identified as that
which maximizes the entropy (simplicity) of the disk maps.  The
determination of a self-consistent distance to HT~Cas was hence one of
the objectives of this study.


\section{The PPEM analysis}

We used the two-parameter spectral model described in VHH99 to map the
temperature $T$ and surface density $\Sigma$ of the disc material.
The assumed spectrum is that of a simple uniform slab,
\begin{equation}
\label{inu}
I_\nu = B_\nu(T) (1 - e^{-\tau_\nu/\cos i})
\end{equation}
where $\tau_\nu = \rho\, \kappa_\nu\, 2 H$ is the optical depth, 
$\rho = \Sigma\, /\, 2 H$ is the mass-density,
$H \approx c_s/\Omega_{Kepler}$ is the vertical scale height
of the slab, here assumed to be the disc's semi-thickness,
$\kappa$ is the mass absorption coefficient,
$c_s$ is the sound speed and
$\Omega_{Kepler}$ the Keplerian angular velocity of the disc material.

Although we
assume a very simple pure hydrogen slab in local thermodynamic
equilibrium (LTE) with no line emission, the difficulty in computing
accretion disc spectra and our ignorance of the true vertical
structure of optically thick or thin discs makes this model still very
useful as a means of distinguishing between optically thick and thin
parts of the disc.

At the densities and temperatures considered here, LTE is a very good
approximation.  The assumption of pure hydrogen is quite adequate for
hot or warm regions of the disc.  For realistic slabs (i.e. including
metals) at temperatures lower than about 6\,300K, the opacity is still
mostly provided by H$^-$ and hydrogen bound-free emission but the
total opacity is much higher due to the increase in the number of free
electrons relative to a pure-hydrogen gas.  Thus, our method {\it
severely underestimates} the intensity emerging from {\it cooler} parts of
the disc.  If a certain flux is required, the algorithm will be forced
to increase the fitted temperature (which increases the opacity in
this range) and/or increase the surface density.  {\it Regions of the
reconstructed disc with kinetic temperatures less than 6\,300K will thus
-- in fact -- have true temperatures and/or surface densities which are
significantly lower than those reconstructed using our simple model}.
Fortunately, this effect will only be important in the
least-constrained outer disc where our ability to reconstruct physical
parameters is the worst anyway (due to the disappearance of the Balmer
Jump and low intensities in the observed filters).  We will see that
our upper limits on the parameters in some
regions are still physically useful.

For theoretical spectra with no line
emission, the physical parameter maps derived from the UBVR magnitudes
are similar no matter whether one convolves the spectra with the
filter band-passes or uses the monochromatic fluxes at the mean
wavelengths of the pass-bands (indeed in the latter case our numerical
results are slightly better). This is due to the fact that the
continuum is very flat (similar fluxes in all four passbands).  The
inclusion of emission lines would improve the sensitivity of the maps
to the U-band fluxes and regions with substantial H$\alpha$, and hence
to the optically thin parts of the disc.  However, this effect goes in
the same direction as that mentioned before for the continuum: these
less realistic models have {\it less} flux in the filter pass-band and
so require higher temperatures and surface densities.

Since HT~Cas was in quiescence at the time of observations, we assumed
a geometrically thin disc for our PPEM study -- an assumption which we
will later verify.

We model the white dwarf emission using theoretical white dwarf
spectra and treat the star as a true spherical star which is occulted
by the secondary, occults the accretion disc behind itself, and can be
preset to be partially occulted by the accretion disc.  Wood \&
Horne~(1990) find that the white dwarf is not likely to be partially
occulted, although they could not completely rule out a partial
occultation of the white dwarf.  Our reconstructions
(Section~\ref{maps}) imply that the inner disc is optically thick and
therefore would lead to an occulted ``lower'' hemisphere of the white
dwarf. On the other hand, evaporation of the inner disc (Meyer \&
Meyer-Hofmeister 1994) could lead to a small hole in the accretion
disc which, with our limited spatial resolution, would be difficult to
detect even with PPEM.  However, for the white dwarf to be completely
visible, the hole must be larger than $r_{\mbox{\SC hole}} = R_{wd}
\tan i = 6.3 R_{wd} = 0.17\Rl$, which should be easily detectable in
the reconstruction. Thus, following Wood \& Horne (1990), we first
chose a fully visible white dwarf but then also reconstructed the
accretion disc using a partially white dwarf for comparison (see
Appendix~\ref{app_wd4}).

We assume the white dwarf has a uniform temperature $T_{wd}$, with no
limb darkening ($u=0$, Wood \& Horne).  Therefore, we use HWS91's white
dwarf radius of 0.0263$\Rl$, which is derived for a vanishing
limb-darkening coefficient. Since the mean wavelengths of the filter
pass-band partly coincide with the absorption lines in the theoretical
white dwarf spectra, we integrated these spectra over the photometric
filter pass-band functions to derive the theoretical white dwarf
flux. These were then used within PPEM to determine the white dwarf
temperature.

The slow ellipsoidal variations of the secondary star during quiescence
and over the small range of orbital phases considered is roughly
$\sim$13\% of the secondary's flux (Marsh 1990): given that the secondary
star contributes certainly no more than 10\% of the light in the R-band
and much less in the others, we neglect the variation in the contribution
of the secondary over the limited phase range between -0.1 and 0.1 and
treat the secondary's light as simply part of the fitted uneclipsed component.

For our reconstructions we used a default map with {\em medium}
blurring, as described in Appendix~\ref{app_default}.

\begin{figure}
\hspace*{0.5cm}
\psfig{file=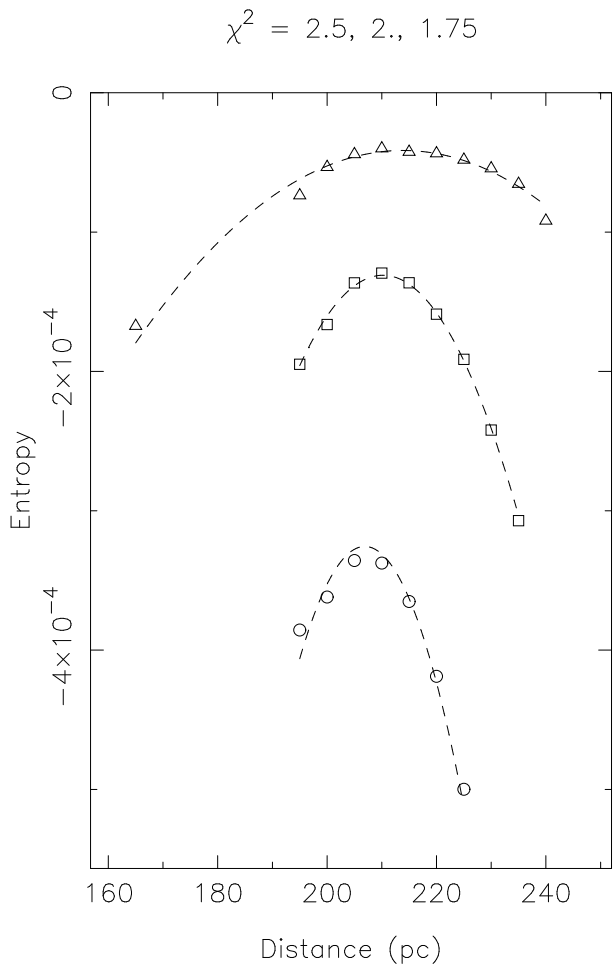,width=8cm}
\caption{\small The entropy of the reconstructions (a measure of
smoothness) as a function of the assumed distance. For each trial
distance the data were fitted with $\chi^2/N = 2.5$ (triangles), 2.0
(squares) and 1.75 (circles). The dashed lines are parabolic fits to
the data, they peak at 214~pc, 210~pc and 207~pc, respectively.
\label{dist}}
\end{figure}

\section{Reconstructions using the $d \sim 150$~pc estimates}

We first attempted to obtain PPEM paramater maps of the disc in HT~Cas
using the short distances of $d = 140-165$~pc estimated by Wood et al.\ (1995)
and Marsh (1990).

Unlike the results of Wood et al., our reconstructions must fit all of
the data simultaneously and so suffer less from additional arteficial spatial
smearing.  Using $d = 165$~pc we see typical features in the maps
arising when too small a distance was chosen (Appendix~\ref{app_short}):
they show emission in two bright regions along an axis perpendicular to the
binary axis with one of them showing a curiously hot pixel.  This is
the algorithm's desperate attempt to preserve the width of the eclipse
profiles while minimizing the emitting area.  The reconstructed disc
using $d \sim 150$~pc would simply produce too much emission if it were
axially symmetric.

We could fit the data with a reduced $\chi^2/N$ of 2.5 using a distance
of 165~pc, but the entropy of the solution is very low compared to the
entropies of the maps corresponding to larger trial distances, i.e.\ there
are more arteficial structures and poorer fits for $d \sim 150$~pc.
Reconstructions using Marsh's smaller distance are even worse.

Otherwise, the reconstructions show similar radial temperature, surface
density and derived parameter profiles as Wood et al. (e.g.\ a flat 8\,000~K
temperature profile within 0.17$\Rl$ except for a nearly optically
thick arteficial spot with a temperature of 10\,000~K) and a disc radius
of 0.35$\Rl$. The white dwarf is reconstructed with a temperature of
about 19\,300~K, close to Wood et al.'s $18\,700\pm1\,800$~K.  Note that
this reconstruction contains disc temperatures which should be
reliable (i.e. are larger than 6\,300\,K). The effective temperature is
also close to expected values of about 7\,000~K at small radii. The optical
depths range between 0.1 to 0.2 except near the white dwarf, where they
reache values of $\sim$0.7.

\begin{figure*}
\hspace*{0.5cm}
\psfig{file=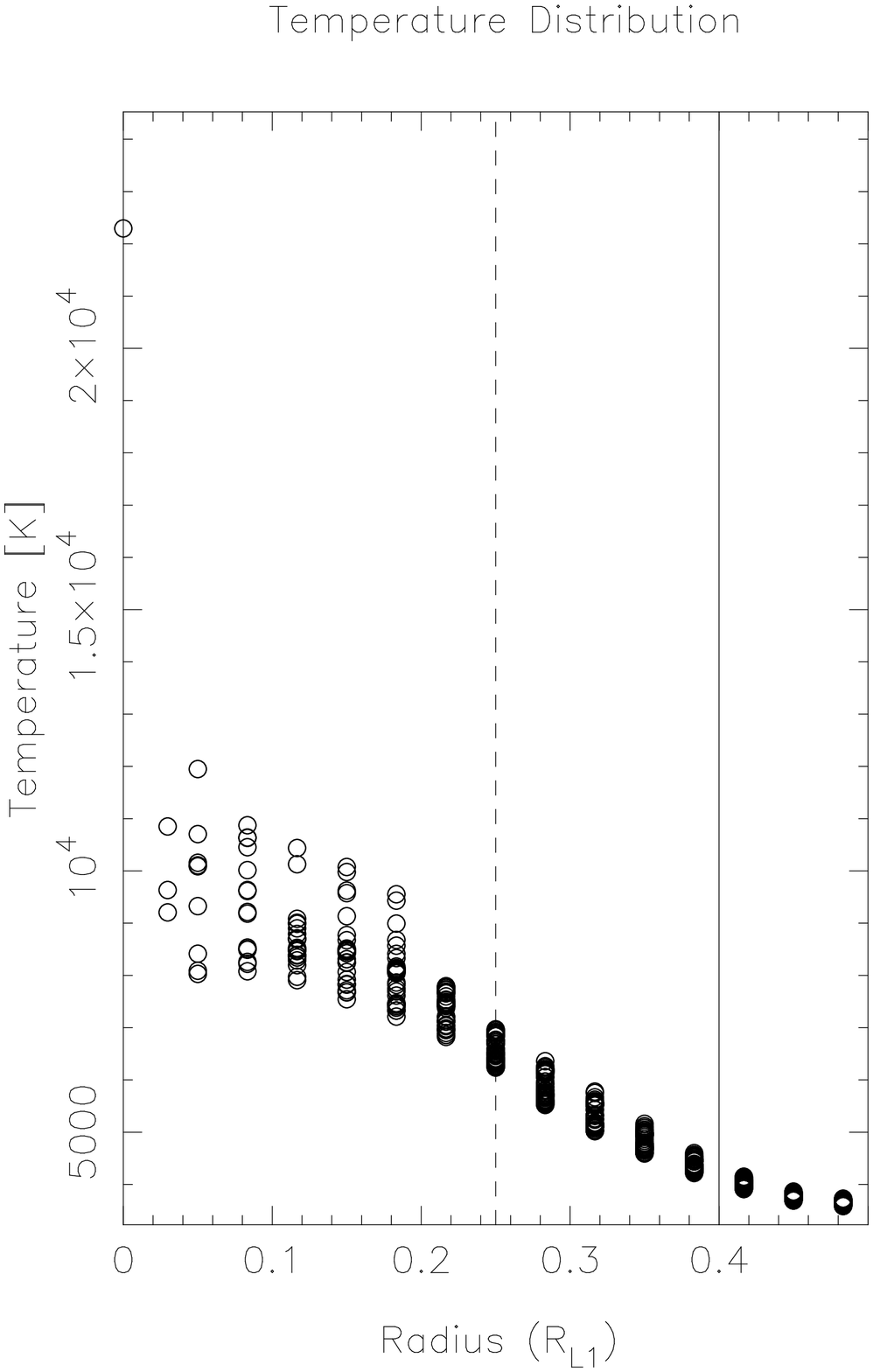,width=7cm}
\hspace*{0.5cm}
\psfig{file=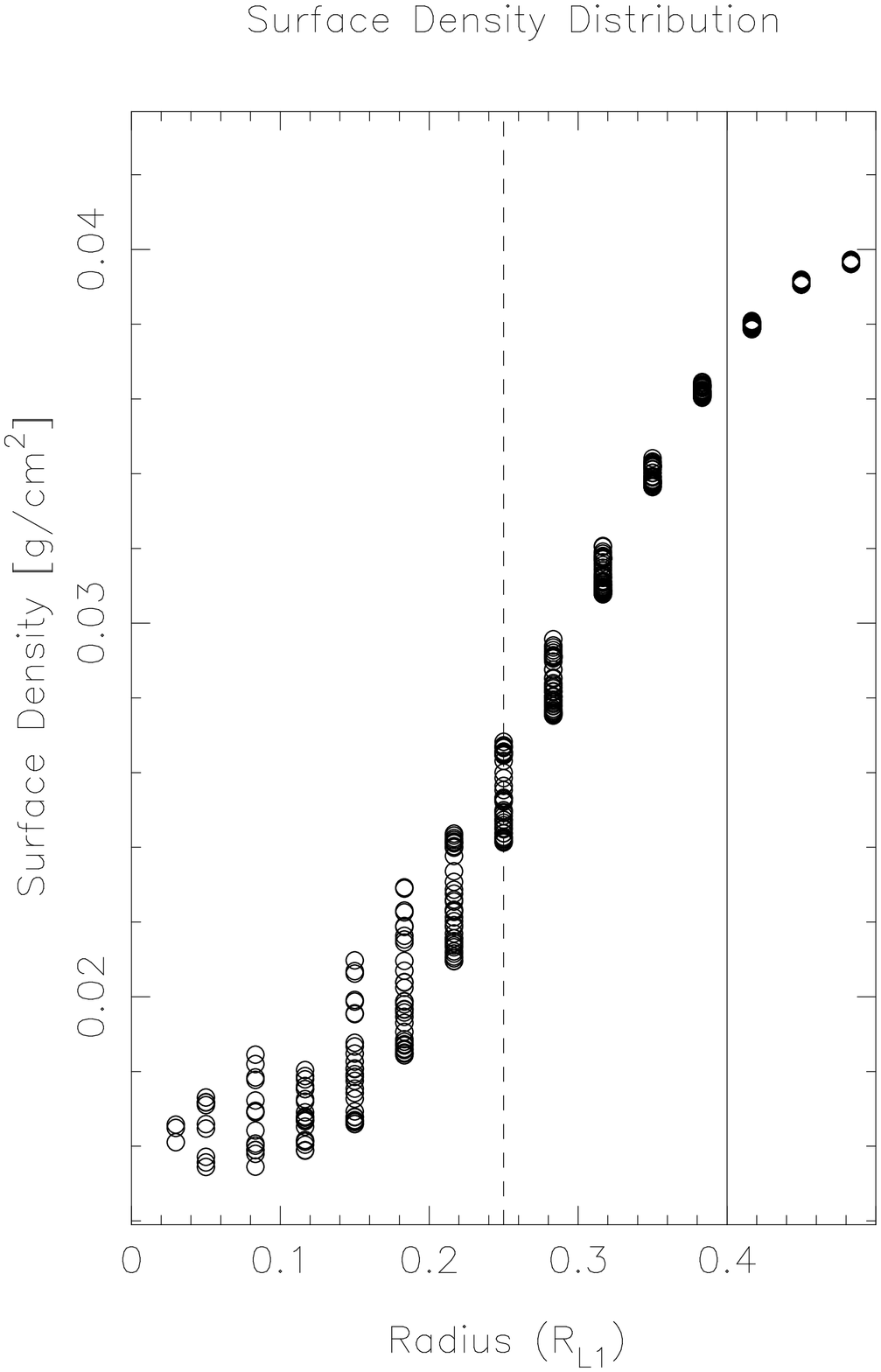,width=7cm}
\caption{\small The reconstructed radial temperature ({\em left}) and
surface density distribution ({\em right}) for a distance of 205pc.
The dashed line at 0.25$\Rl$ indicates that for larger radii the parameter
values become ambiguous, the solid line marks the disc edge.
\label{htp}}
\hspace*{0.5cm}
\psfig{file=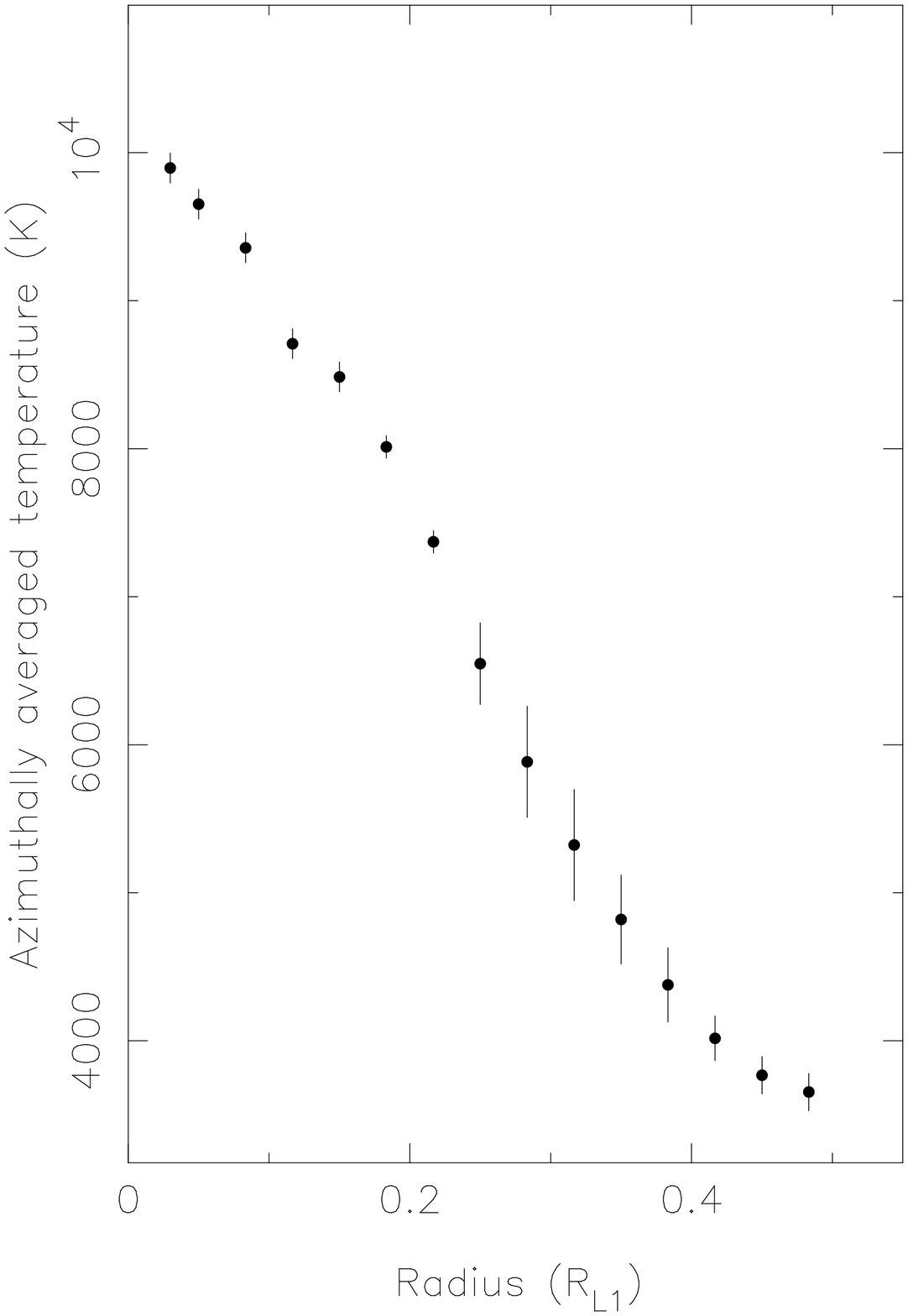,width=7cm}
\hspace*{0.5cm}
\psfig{file=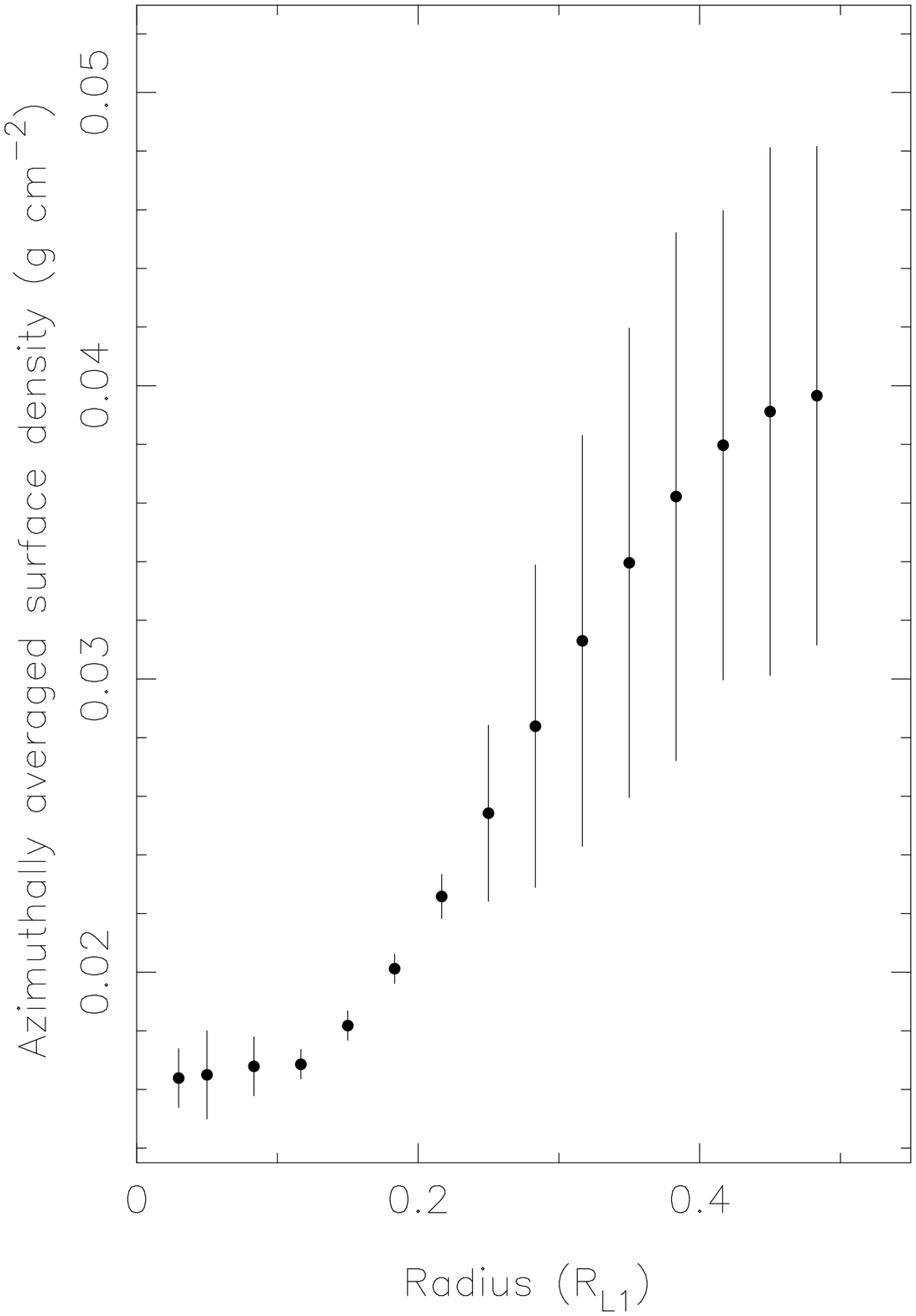,width=7cm}
\caption{\small Azimuthally averaged temperature ({\em left}) and
surface density ({\em right}) distributions with error bars according
to an analysis of the T-$\Sigma$ parameter space.
\label{hta}}
\end{figure*}

\section{Reconstructions including a fitted distance}
\label{distestimate}

Given the problems in the reconstructions for $d \sim 150$~pc
estimates, we next attempted to find the distance to HT~Cas
by using the entropy information in the maps (Vrielmann et al.\ 2002).

We reconstructed the accretion disc maps for a set of trial distances
and determined the entropy for a fixed $\chi^2/N$.  The result is
presented in Fig.\,\ref{dist}.  A parabolic fit to the function for
$\chi^2 = 1.75$ peaks at 207~pc, a distance which is significantly
larger than the previous estimates using the white dwarf or the
secondary flux. As a confirmation for this value, in
Appendix~\ref{app_long} we show the maps using a larger distance of
240~pc showing typical features of too large a trial distance used.

We estimated the error in the PPEM distance estimate by performing
the test described in Appendix~\ref{app_dist} using an arteficial disc
with a dominant bright spot.  The resulting distance is 2.5\% larger
than the original one.  This translates into an error of at least
$\pm$5~pc for our distance of 207~pc.

In the following Sections we present the maps and derived quantities
using the distance of 205~pc which should effectively give the same
results as 207~pc. We use this distance although it does not agree
with previous estimates and although we do accept the smaller distance
of 133~pc because -- as we will explain in Section~\ref{patchydisc} --
the maps do represent part of the real disc.

\subsection{The light curves}
\label{lightcurves}

Despite the simplicity of the hydrogen model, we find a very good fit to
the eclipse light curve, as shown in Fig.\,\ref{htf} with a $\chi^2/N$
of 1.75. The residuals (at the bottom of the plot) show little correlation,
apart from the flickering between phases 0.055 and 0.075.  In the U filter,
this looks like the re-appearance of a bright spot, but the light curves
in the filters V and R rather suggest two flares at phases 0.055 and 0.07.

The residuals between the observed and fitted light curves are at most
5.6\% throughout the eclipse (between phases -0.05 and 0.05), in general
largest for the V light curve and smallest for U and B, i.e.\ the Balmer
Jump is very well reconstructed.  The larger residuals of the V and R light
curves may be due to the lack of line emission in our model, particularly
H$\alpha$. This is also compatible with the R light curve having slightly
negative residuals in mid-eclipse.

\begin{figure*}
\hspace*{0.5cm}
\psfig{file=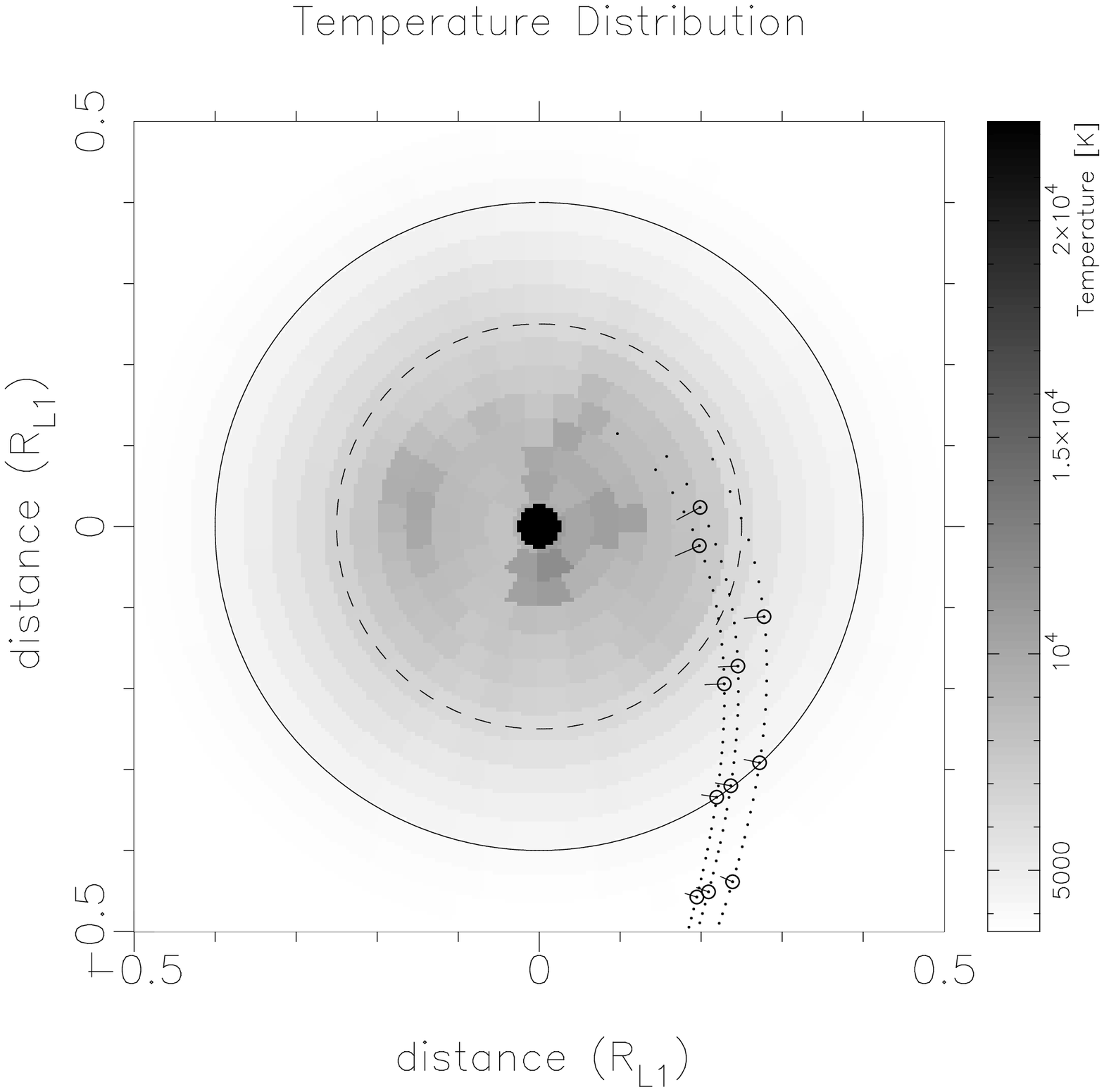,width=8cm}
\hspace*{0.5cm}
\psfig{file=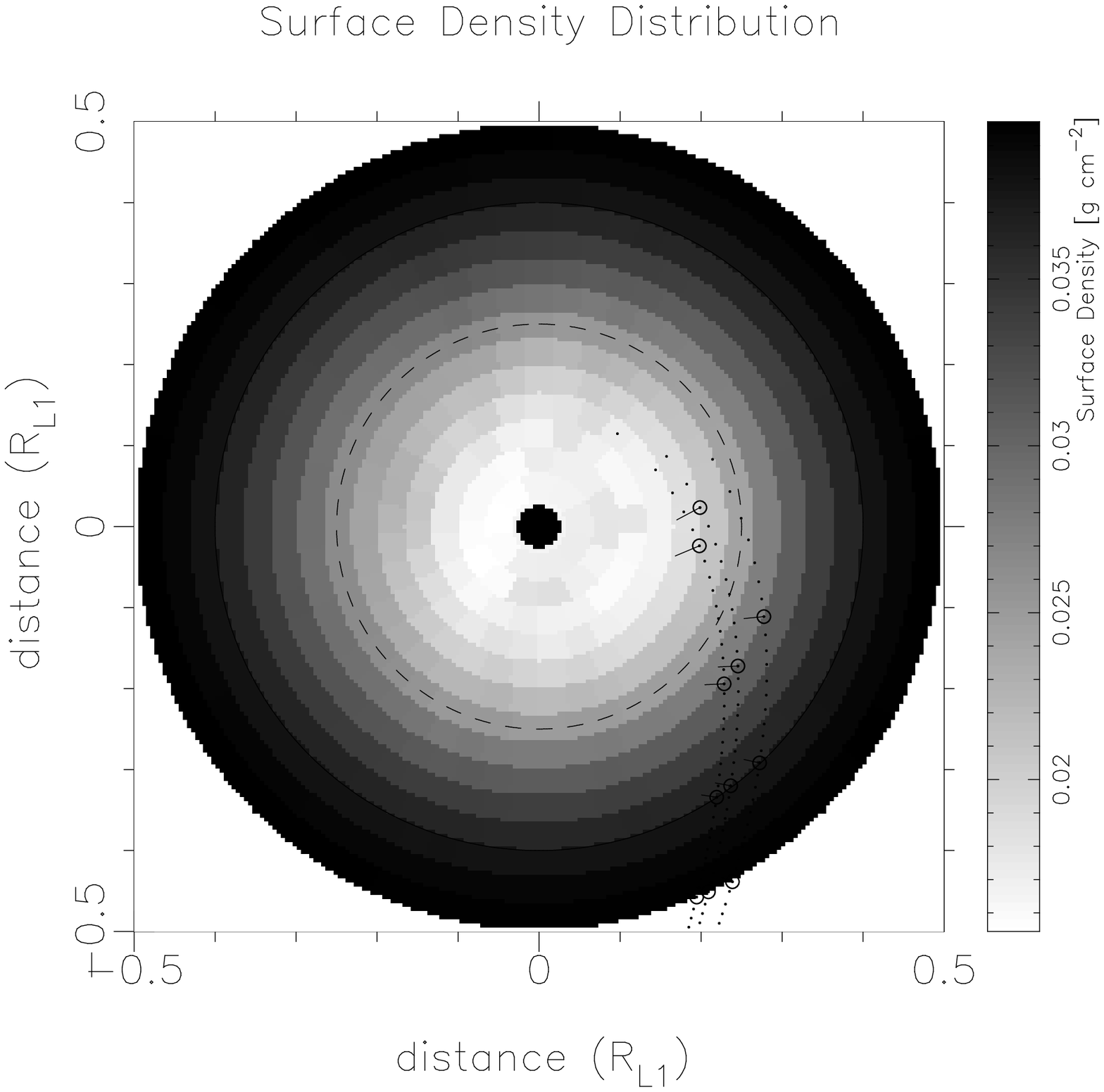,width=8cm}
\caption{\small The reconstructed temperature ({\em left}) and surface
density ({\em right}) map. The Roche-lobe of the primary component lies just about
outside the plotted region. The dashed line at
0.25$\Rl$ indicates that for larger radii the reconstructed parameter
values become ambiguous, the solid line marks the disc edge. The
curved, dotted lines in the grey-scale plot are theoretical accretion
stream paths for mass ratios 0.15 $\pm$ 50\%. The secondary is at the
bottom.
\label{htmaps}}
\end{figure*}

\subsection{The reconstructed accretion disc}
\label{maps}

The reconstructed temperature $T$ and surface density $\Sigma$
distributions (maps) are displayed in Fig.\,\ref{htp} and
Fig.\,\ref{htmaps}. The most prominent features are the radially
decreasing temperature and increasing surface density. However, before
reliable quantitative results can be drawn from these maps, the
($T,\Sigma$) parameter space has to be analysed in a way described in
detail in VHH99.

This analysis shows that both the temperature and surface density
values are very well constrained up to a radius of 0.25$\Rl$.  For
radii larger than 0.25$\Rl$, the fitted temperatures and surface
densities become significantly anti-correlated and hence ambiguous.
Since we reconstruct both parameters simultaneously, we cannot rely on
the exact values of these two parameters for radii larger than
0.25$\Rl$ and any derived parameters (with the exception of the
effective temperature, see Section~\ref{sec_teff}).  To illustrate the
reliability of the reconstructed values of $T$ and $\Sigma$,
Fig.\,\ref{hta} shows azimuthally averaged parameter distributions with
approximate error bars calculated using the assumption that the
spectrum at each particular point may vary by up to a $\chi^2/N$ of 3.
This $\chi^2/N$ is larger than that of the fit to the data and
therefore allows some deviation of the parameters due to MEM smearing.

Given local values of temperature and surface density, we calculate
the specific intensity distributions and find that the inner disc's
R-band intensity is 30\% of that of the white dwarf.  This drops with
radius to 7\% at 0.25 $\Rl$, and to less then 1\% at 0.4 $\Rl$ which
we call the disc edge.  The region outside 0.25 $\Rl$ contributes so
little flux that the data provide only upper limits: the $T$ and
$\Sigma$ values found in this range are controlled by the entropy
extrapolating the radial $T$ and $\Sigma$ gradients defined by the
data for the region inside 0.25 $\Rl$.

The temperature in the inner part of the disc lies at about 9\,500~K
(with a range of 8\,500~K to nearly 11\,000~K) and drops to about 7\,000~K
at 0.25$\Rl$ and to 4\,000~K at the edge of the disc. The surface density
increases in the same radial range from about 0.013\,\gcm\ to about
0.026\,\gcm\ at a radius of 0.25$\Rl$ and 0.038\,\gcm\ at a radius of
0.4$\Rl$.  Given that the temperatures for $R > 0.25 \Rl$ are lower than
6\,300\,K, -- corresponding to temperatures where our spectral model is bound
to underestimate both physical parameters --  the true temperatures and/or
surface densities are likely to be lower.  Thus, we estimate that
the outer disc radius is 0.3-0.4 $\Rl$.

\begin{figure*}
\hspace*{0.2cm}
\psfig{file=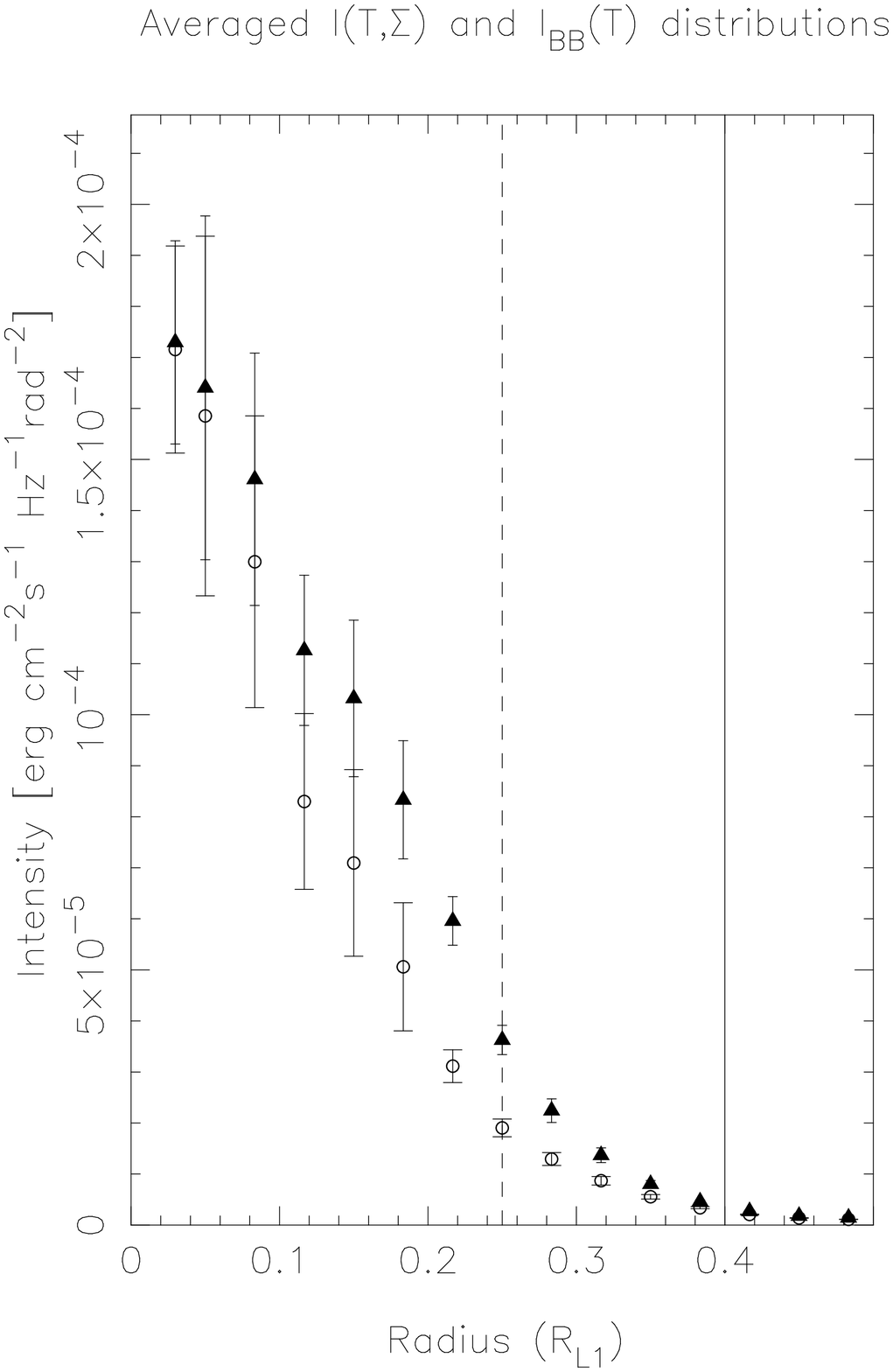,width=6cm}
\hspace*{1cm}
\psfig{file=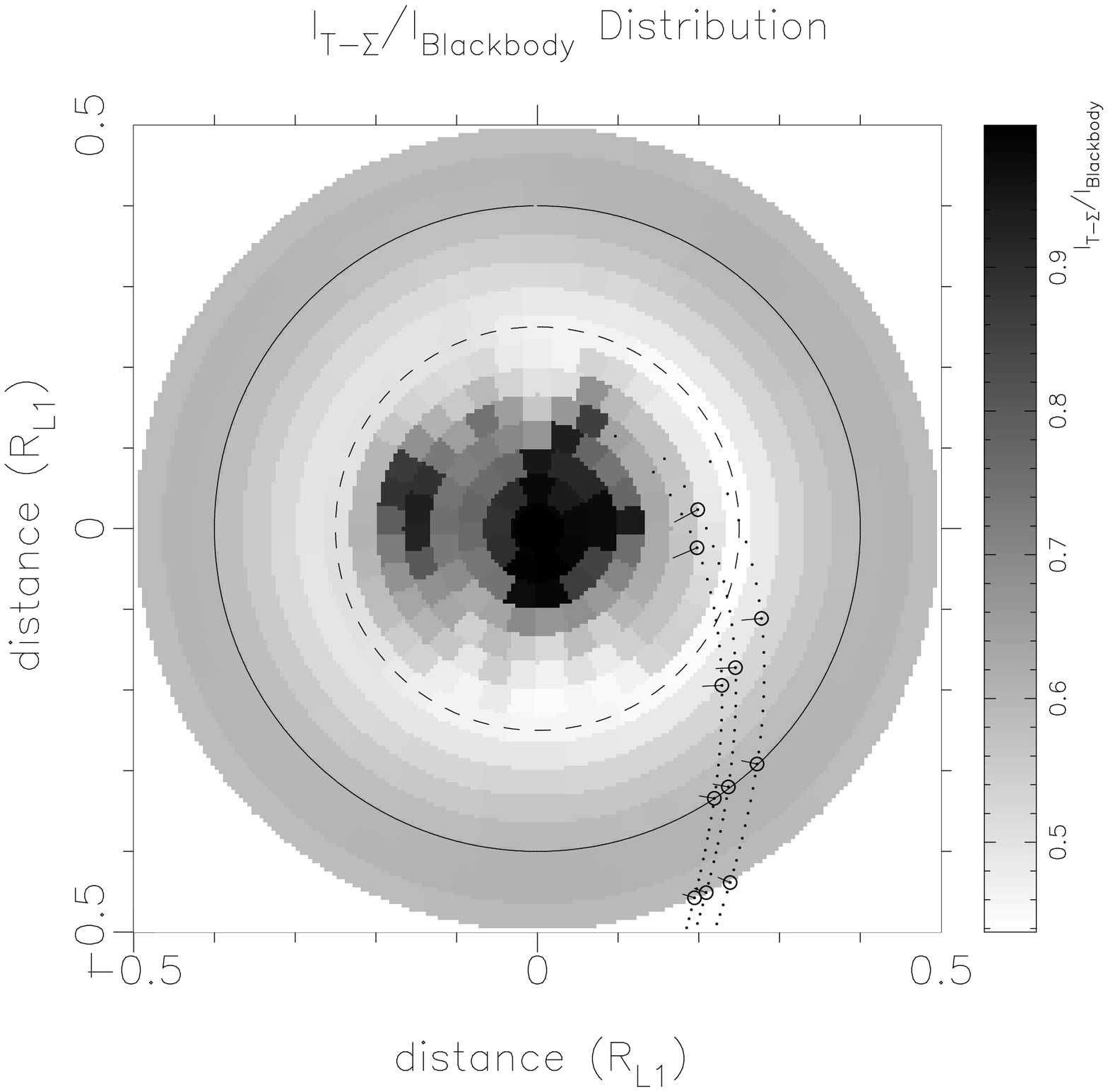,width=9.5cm}
\caption{\small {\em Left:} The radial distributions of the spectrally
and azimuthally averaged intensity distributions of $I(T,\Sigma)$ and
$I_{BB}(T)$. {\em Right:} The distribution of the ratio
$I_{T,\Sigma}/I_{BB}$ as a gray-scale plot (see text for
explanation). The Roche-lobe of the primary component lies just about
outside the plotted region. For both plots: The dashed line at
0.25$\Rl$ indicates that for larger radii the reconstructed parameter
values become ambiguous, the solid line marks the disc edge. The
curved, dotted lines in the grey-scale plot are theoretical accretion
stream paths for mass ratios 0.15 $\pm$ 50\%. The secondary is at the
bottom.
\label{htbb}}
\end{figure*}

\begin{figure}
\psfig{file=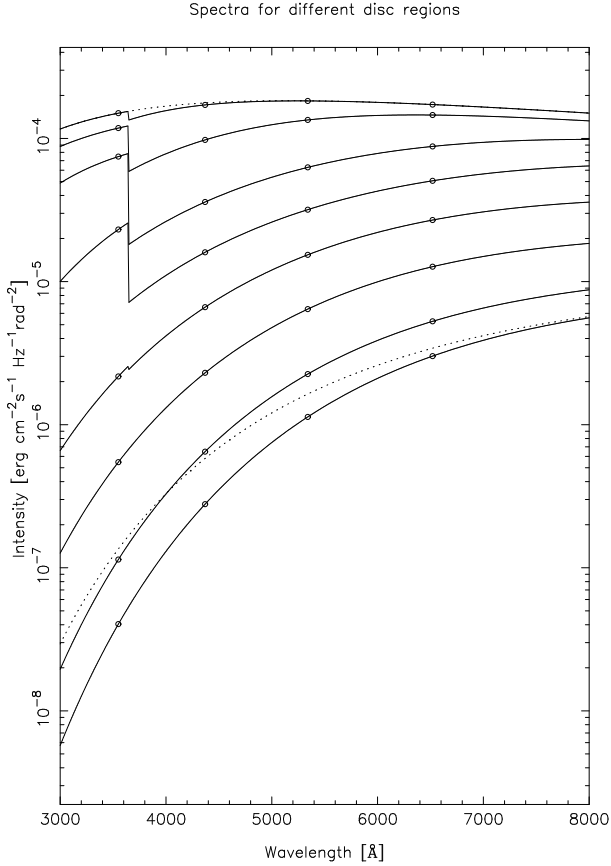,width=8cm}
\caption{\small Reconstructed intensity spectra for disc radii 0.03
(uppermost), 0.08, 0.15, 0.22, 0.28, 0.35, 0.42 and 0.48 $\Rl$
(lowermost) with overall intensity decreasing monotonically with
increasing radius.  The parameters $T, \Sigma$ used to calculate the
spectra are the averaged values from Fig.\,\ref{hta}.  Additionally,
black body spectra are shown for comparison at temperatures 9\,900~K
(upper dashed line) and 3\,650~K (lower dashed line), corresponding to
the inner and outer disc temperatures at 0.03 and 0.48 $\Rl$,
respectively; the circles indicate the mean wavelengths of the UBVR
band-passes.
\label{spec}}
\end{figure}

To obtain a bolometric measure of the optical depth, we compare
the spectrally averaged intensity distribution $I(T,\Sigma)$ to the
black body intensity distribution $I_{BB}(T)$ derived from the
reconstructed temperature $T$ (Fig.\,\ref{htbb}).  The ratios of
the intensities at each radius show that all of the disc has an
optical depth of order unity, but the inner parts are more nearly
optically thick.
A more quantitative picture is drawn by the spectra of different disc
regions as shown in Fig.\,\ref{spec}. The parameters are selected according
to the azimuthally averaged maps (Fig.\,\ref{hta}). While the Balmer Jump is
present in the central parts, it disappears for radii larger than
0.25$\Rl$, because of the low intensity at the blue end of the
spectrum. Note that the disc is optically thin at all radii in the
B-band, is marginally thin in the V-band, is optically thick in the
inner disc in the U-band, and nearly optically thick in the R-band.

The leading side of the inner disc has higher optical depths and so is
closer to a black-body radiator. This coincides with the larger
temperature and surface density values seen in Fig.~\ref{htp} at small
radii. This asymmetry is reproduced in the eclipse profile of the
accretion disc.  To explain the asymmetry in the light curve, we
seperate the phases in four different ranges $A,B,C,D$ with $A <
\phi_{wd,i}$, $\phi_{wd,i} < B < 0$, $0 < C < \phi_{wd,e}$ and
$\phi_{wd,e} < D$ where $\phi_{wd,i}$ and $\phi_{wd,e}$ denote the
phases of white dwarf ingress and egress, respectively.  Before the
white dwarf is eclipsed, we see a gradual, linearly sloped disc
ingress (phase range $A$). While the central star is obscured, the
disc ingress profile is short and rounded, reaching minimum light
before phase 0 ($B$). The following egress ($C$) is again more gradual
and after the reappearance of the white dwarf we see the short, steep
egress of the remaining disc ($D$). Phases $A$ and $C$ with the more
gradual profiles correspond to the following lune, phases $B$ and $D$
to the leading, brighter lune of the accretion disc.  Hence, the
leading lune has a steeper gradient in the parameter distributions
than the following lune. A comparison with a much more symmetric
reconstruction of the disc corresponding to a fit to the data with
$\chi^2/N = 2.5$ shows that exactly these features in the eclipse
profile are responsible for the asymmetric parameter distributions. At
this higher $\chi^2/N$ level we see correlated residuals especially
inside the white dwarf eclipse profile.

For comparison, we reconstructed the accretion disc using a partially
visible white dwarf. The results are described in Appendix~\ref{app_wd4}. We note
here only that considering the PPEM method, the solution with the
partially visible white dwarf is inferiour to the above presented solution.

\subsection{The white dwarf}
\label{sec_wd}

Our white dwarf fluxes are 0.55~mJy, 0.51~mJy, 0.43~mJy and 0.31~mJy
for the filters UBVR, respectively. Except for the U flux, our white
dwarf fluxes are smaller than WHS's fluxes (their Table 1). The
remaining flux in the BVR filters must originate in the inner part of
the accretion disc.

The reconstructed temperature of the white dwarf is 22\,600~K. This
value is larger than Wood et al.'s (1995) value of $18\,700\pm1\,800$~K
determined from the same data, because of the larger distance used.
If we use a partially obscured white dwarf (Appendix~\ref{app_wd4}) we
arrive at a white dwarf temperature of $T_{wd,1/4} = 26\,700$K, even
larger than Wood et al.'s (1995) value.

The fact that PPEM favours a fully visible white dwarf is problematic,
because this means we should expect a substantial hole in the inner part
of the disc of the radial size $0.17\Rl$. Such a hole is not evident
in our reconstructions. However, Wood \& Horne's (1990) analysis of the
white dwarf ingress and egress profiles also favours a fully
visible white dwarf or a white dwarf fully covered by a boundary
layer, but specifically excludes a white dwarf with $u=0$ and an
occulted lower hemisphere.

In our study we have not specified a location of boundary layer, but
expected it to be reconstructed in the inner disc. Since we do not see
any indication of a boundary layer in the disc, its temperature must
be mixed into the white dwarf temperature, making it in our model
indistinguishable from the central object. In this scenario might lie
the solution to the problem, since the boundary layer has a different
geometrical and radiative structure than a white dwarf.

The white dwarf or boundary layer appears to have different
temperatures at different times, as reported by Wood et al. (1995): it
varies between 13\,000 and 20\,000~K, lower temperatures associated
with low states of the accretion disc. The white dwarf and/or boundary
layer cools down during quiescence when the accretion rate is reduced
and is heated up again during an outburst.  Since the HT\,Cas data
were taken 2-3 years after the previous outburst, the temperatures
derived using $d \sim 200$~pc would appear to be too high compared
with those of white dwarfs in similar dwarf novae.  However, the
current mean mass-accretion rate does not necessarily reflect the
long-term rate, so that high temperatures could principally be due to
a previous phase of high mass-accretion rates (e.g. Schreiber,
G\"ansicke \& Cannizzo 2000).

\subsection{The uneclipsed component}
\label{uneclipsed}

\begin{figure}
\hspace*{0.2cm}
\psfig{file=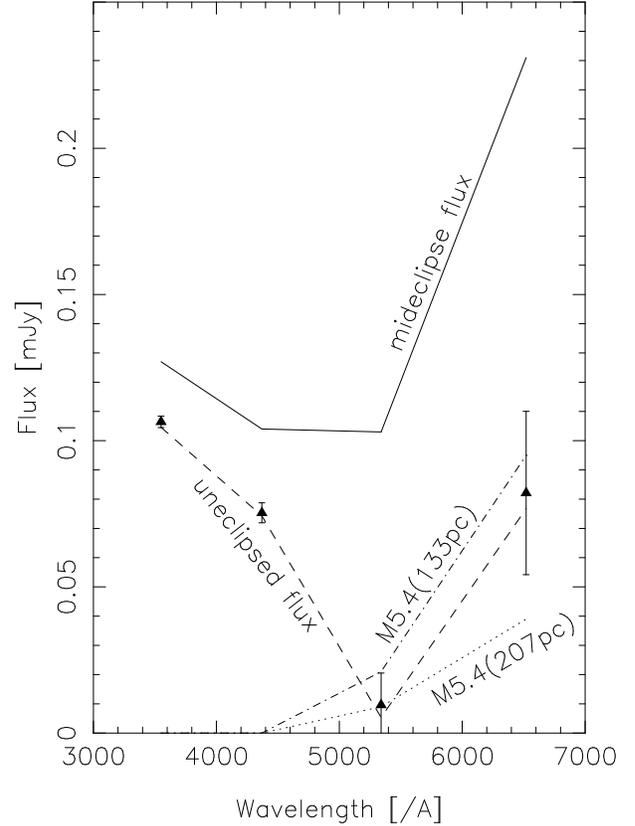,width=8cm}
\caption{\small The uneclipsed component (dashed line) in comparison
to the mideclipse flux (solid line) and the contribution of a M5.4
dwarf star at a distance of 133~pc (dash-dotted line) and at 207~pc
(dotted line). The points with error bars give a mean and standard
deviation of the uneclipsed flux for reconstructions with three
different default maps ({\em little}, {\em medium} and {\em much}, see
Appendix~\ref{app_default}) to illustrate the uncertaincy of the
reconstructed value.
\label{unecl}}
\end{figure}

\begin{table*}
\caption{The reconstructed fluxes in mJy of the uneclipsed component
($F_{\mbox{\SC un}}$), the total mid-eclipse fluxes ($F_{\mbox{\SC
mid}}$), and the the total out-of-eclipse fluxes $F_{\mbox{\SC out}}$
compared with the flux of an M5.4 main sequence star at 133 and
207\,pc ($F({\mbox{\SC M5.4}})$). Columns 5 and 6 give the mean
and standard deviation of the uneclipsed flux for reconstructions with
three different default maps (little, medium and much, see
Appendix~\ref{app_default}) to illustrate the uncertaincy of the
reconstructed value.
\label{tab_ucq}}
\vspace{1ex}
\hspace{0.5cm}
\begin{tabular}{cccccccc}
	Filter &
        $F_{\mbox{\SC out}}$ &
        $F_{\mbox{\SC mid}}$ &
        $F_{\mbox{\SC un}}$ &
	$<F_{\mbox{\SC un}}>$ &
	$\sigma(<F_{\mbox{\SC un}}>)$ &
        $F_{\mbox{\SC M5.4}(133)}$ &
        $F_{\mbox{\SC M5.4}(207)}$ \\ \hline
 U  & 1.13 & 0.127 & 0.099 & 0.106 & 0.002 &   --  &   --  \\
 B  & 0.98 & 0.104 & 0.078 & 0.075 & 0.003 &   --  &   --  \\
 V  & 1.00 & 0.103 & 0.018 & 0.010 & 0.011 & 0.021 & 0.009 \\
 R  & 1.20 & 0.231 & 0.124 & 0.082 & 0.028 & 0.095 & 0.039 \\
\hline
\end{tabular}
\end{table*}

The uneclipsed component is the fitted flux in each band-pass which does
not appear to be eclipsed and includes the contribution of the secondary
star.  The uneclipsed component comprises no more than 8-10\% of the total.
It increases to the UV, indicating a hot uneclipsed source like a chromosphere
above the disc (Fig.\,\ref{unecl}), reaching so far above the disc that it is
never eclipsed (i.e.\ a few white dwarf radii).

Assuming an M5.4 main sequence star (Marsh 1990) and using the Kirkpatrick
\& McCarthy (1994) flux tables, we can estimate the contribution of
the secondary star to the uneclipsed component (Tab.~\ref{tab_ucq}).
{\it All} the uneclipsed light in V and R can be accounted for
if we assume the smaller distance of 133~pc to be the correct
one (see Tab.~\ref{tab_ucq}). The fact that actually the expected V
and R fluxes are somewhat higher than the reconstructed one in this case
is due to the uncertainty in the determination of the uneclipsed flux. However,
it could also indicate that the secondary is more likely of slightly
later type, a suggestion mentioned in Section~\ref{distance}.
Alternatively, the hot uneclipsed source is bound to have some V and R
flux, a situation easily accounted for in the solution with the larger
PPEM distance in which the secondary contributes less light.

\begin{figure}
\psfig{file=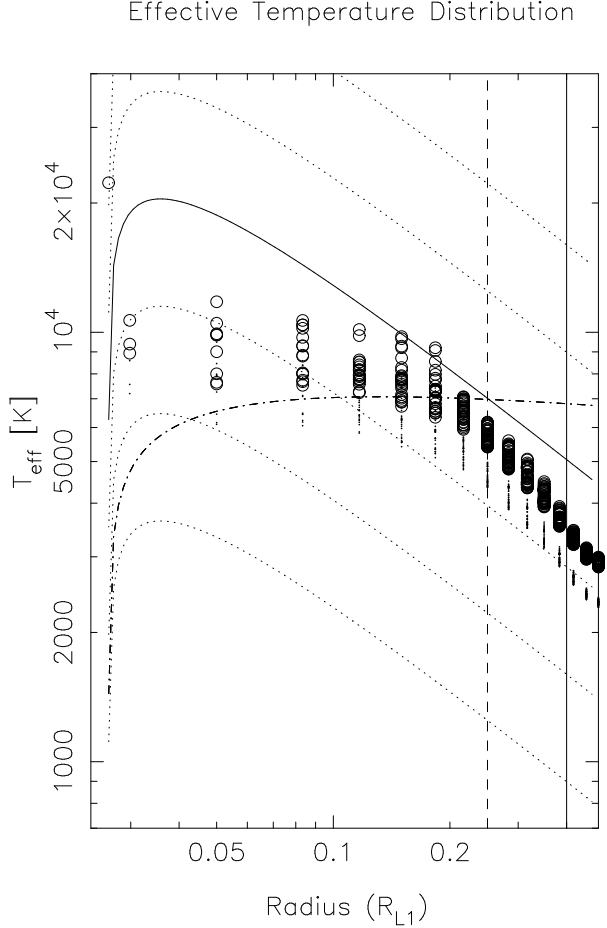,width=8cm}
\caption{\small The radial effective temperature distribution for a
distance of 205\,pc (circles) and 133\,pc (dots) assuming a
covering factor of 41\%. The underlying dotted lines are theoretical
steady state temperature distributions for $\log \Md = 13$ to 18, the
one for $\Md = 10^{16}$gs$^{-1}$ is drawn solid for reference. The
dashed-dotted line indicates the critical mass accretion rates
\protect{$\dot{M}_{\rm A}$} according to Ludwig et al.\ (1994). The
vertical dashed line at 0.25$\Rl$ indicates the radius above which the
reconstructed parameter values become ambiguous. The vertical solid
line marks the disc edge.
\label{htteff}}
\end{figure}

\subsection{Derived parameters}

The reconstructed T-$\Sigma$-maps allow us to derive further parameters like
the effective temperature or mass-accretion rate distribution,
the apparent viscosity, and the disc height and mass.

\label{sec_teff}

\vspace{1ex}
\centerline{\it The effective temperature $\Teff$}

\vspace{1ex}
The effective temperature is calculated as the integral of the
intensity of the emission over all directions $\Omega$ and frequencies $\nu$
\begin{equation}
\int\!\!\!\int I_\nu(T,\Sigma) \,\,d\nu \,\, d\Omega = \sigma \Teff^4
\label{eq_teff}
\end{equation}
with $\sigma$ the Stefan-Boltzmann constant and $I_\nu$ according to
Eq.~\ref{inu}.

Fig.\,\ref{htteff} shows the effective temperature distribution
calculated directly from our spectral model and the fitted parameters.
Even though the reconstructed parameters $T$ and $\Sigma$ are somewhat
ambiguous for radii larger than $0.25 \Rl$, the effective temperature
$\Teff$ is reasonably reliable, since I($T$, $\Sigma$) must
still be reproduced (the light curves are fitted very well) and the
optical flux is a large fraction of the bolometric flux.

Within a radius of 0.2$\Rl$ the $\Teff$ distribution is relatively
flat, with values around 7\,500~K or 10\,000~K. The
distribution of $\Teff$ is distinctly spatially separated, following
the pattern of the optical depth (with higher $\Teff$'s corresponding
to the optically thick and lower $\Teff$'s to optically thin regions).
For radii larger than 0.2$\Rl$, the disc appears to be compatible with
a steady state disc with a mass accretion rate around
$3\times10^{15}$gs$^{-1}$ (i.e.\ $5\times10^{-11}\Myr$).

Also shown in Fig.\,\ref{htteff} is the critical effective temperatures
as a function of radius predicted by Ludwig et al.\ (1994). This
corresponds to the critial mass-accretion rates above which we would
expect steady accretion in an optically thick disc to occur.  Within a
radius of 0.2$\Rl$ our derived $\Teff$ lie above this line. Since the
data were taken in the middle of a quiescent period, the effective
temperatures should not have reached above the critical values.

\vspace{1ex}
\centerline{\it The viscosity}

\vspace{1ex} The most important parameter describing the disc is the
viscosity $\nu$, usually parametrizised as $\nu = \alpha c_s H$
(where $c_s$ is the local sound speed and $H$ the disc scale height)
using the Shakura \& Sunyaev (1973) $\alpha$-parameter.  In order to
be able to fit the dwarf nova light curves with current accretion disc
models, we need values of $\alpha$ around 0.01 in quiescence (e.g.\
Smak 1984).

Using the standard relation between the viscously dissipated and
total radiated flux $F_\nu = I_\nu d\Omega$,
\begin{equation}
\label{viscosity}
2 H \alpha P \frac{3}{4} \Omega_K = \int_0^\infty F_\nu d \nu = \sigma \Teff^4
\end{equation}
where $P$ is the (gas) pressure, we can map the viscosity parameter
$\alpha$ of the disc material.  The derived values lie between 20 and
230, far larger than those expected in any kind of turbulent accretion
disc and similar to those derived by WHV92.

\vspace{1ex}
\centerline{\it The disc height and mass}

\vspace{1ex}
The equilibrium disc height ratio, $H/R = c_s /
\Omega_{\mbox{\SC K}}$ ($c_s$ is the local sound speed and
$\Omega_{\mbox{\SC K}}$ the Keplerian angular velocity of the disc
material) is roughly 0.01, i.e.\ the disc has a very small opening
angle of a degree or less.  The innermost part of the disc and the
disc towards the edge are even narrower (with an opening angle of
approximately $0\gdot4$) giving the disc a slightly concave shape in
the inner region and convex shape in the outer part. In any case, the
reconstructed disc is extremely thin, justifying (at least internally)
our thin disc approximation.

Within a radius of 0.4$\Rl$ the mass of the disc is $5.4
\times 10^{19}$~g ($2.7 \times 10^{-14} {\cal M}_\odot$). This factor
is uncertain by up to 50\% due to the ambiguity of the surface density
density values in the outer parts. A more robust calculation of the
disc mass for the inner part of the disc (up to a radius of 0.25$\Rl$)
leads to $1.3\times 10^{19}$~g ($6.4\times 10^{-15} {\cal M}_\odot$).


\section{The resolution of the discrepancy}
\label{resolution}

Our PPEM distance of 207\,pc is significantly larger than the
estimates of $140\pm 14$\,pc by Marsh (1990), $165\pm 10$\,pc by
Wood et al.\ (1995) and our revised estimate of $133\pm14$\,pc from
Section \ref{distance}.  The main problem with the reconstructions
using $d \sim 150$~pc is the appearance of characteristic
distortions due to too much flux from the nearer disc, a problem
which can easily be remedied by increasing the distance.  On the
other hand, the PPEM reconstructions at $d \sim 200$~pc also have
their problems, notably effective temperatures, $\Teff$, which are
higher than those expected for quiescent discs.  {\it Both} solutions
suffer from the fact that the values of the viscosity paratmers,
$\alpha$ are orders of magnitude higher than can be physically explained.

We therefore need to check quite critically the
potential pit-falls in our own analysis and -- if the method can be
shown to be reliable enough -- to explain the behaviors of the
different reconstructions using an improved physical model of the disc.

\subsection{Reliability of the PPEM method}

If our reconstructions would like to have a distance which is too high,
we must identify which false assumption has produced
this result.  There are several possible explanations for the discrepancy.
\begin{itemize}
 \item This effect could be a peculiar side-effect of the PPEM approach
	since it did not show up in classical eclipse mapping analyses.
	However, the latter do not use the constraints on the colors and
	spectral shapes of the mapped regions and are thus free to yield
	even very unphysical solutions.  The simultaneous use of all the
	information as well as the physical constraints posed by the
	physical (if simple) spectral model should make PPEM more
	sensitive to such effects. Thus, it is not surprising that this
	problem did not arise in WHV92s analysis and there is no a priori
	reason why PPEM should not produce reasonable results.

 \item Wood et al.'s distance is based on the white dwarf colours alone,
 	which were deduced from the original eclipse light curve while
	our analysis includes the information of the accretion
	disc. Furthermore, the distance Wood et al. fitted is not
	independent from the temperature fitted to the white dwarf
	fluxes. We believe our {\em method} is therefore more advantageous.

 \item As we have explained, our model for the emission from the disc does
	not necessarily describe the true emissivity of the disc, both
	because we have no a priori knowledge of the structure of the
	disc and because of our use of pure hydrogen spectra.
	However, the reason that our algorithm produces a larger
	distance is that the surface brightness of the disc is
	basically set by the colours and the eclipse width, which, for
	a given orbital separation, corresponds to a maximum size of
	the emitting region.  Solutions with smaller distances have
	disc fluxes which are too large for the given spectra and
	areas, i.e.\ the specific intensity of the disc would have to
	be {\it smaller}. Lowering the surface density would decrease
	the intensities, but also changes the colours, since this
	would produce a large change in the optically thinnest part of
	the spectrum.  The algorithm therefore reacts by
	reconstructing a disc with much more scatter, i.e.\ a much
	smaller entropy and a more or less elliptically shaped
	emission region with the long axis perpendicular to the binary
	axis. This is an unphysical response to a physically constrained but
	primarily numerical problem.  Our assumption of pure hydrogen
	spectra already {\it underestimates} -- not {\it
	overestimates} -- the intensities from the cooler outer disc.
	Thus, we expect that the use of improved spectral models would not
	necessarily solve this distance problem.

\item	The distance could also be affected by our assumption of a pure
	hydrogen spectral model neglecting any line emission from
	hydrogen and other elements. While it is formally not
	difficult to include the lines, the theoretical line fluxes
	are more dependent upon the detailed model of the vertical
	structure of the disc, are principally more sensitive to the
	diffuse outer layers of the emission region which do not
	contribute much to the total optical depth, and suffer from
	direction-dependent opacity effects due to the shear in the
	disc (Horne \& Marsh 1986). Only a test with an apropriatly
	complicated model would clarify this; in a future paper we
	will use more realistic disc spectra calculated with models
	from Hubeny (1991).  A preliminary analysis is shown in
	Vrielmann, Still \& Horne (2001). We previously noted, that
	when hydrogen dominates the continuum opacity and supplies
	most of the electrons, then metals have little effect on the
	spectrum. Only in cool disc regions will the metals have a
	noticable effect on the opacity. However, these regions have a
	low intensity. We therefore think that our assumption is
	acceptable and that the lines do not play a major role at the
	high densities implied by the reconstructed maps.

\item	Another assumption that might influence the distance estimate is
	our neglect of a dramatic vertical stratification of the disc.
	While the effect on the distance determination is also difficult to
	estimate and a full consideration lies beyond the scope of
	this paper, we consider a scenario in which the disc consists
	of a cool disc with a hot chromosphere. In this case, the
	main emission source is at the base of the chromosphere.
	Because chromospheres tend to stabilize around 10,000~K due to
	the strong dependence of opacity on temperature when hydrogen is
	ionized (Williams 1981, Tylenda 1981),  our assumption should
	not be so bad.

\item 	And finally, Vrielmann (2001) \& Vrielmann et al.~(2002) were
	able to determine a PPEM distance for UU~Aqr and V2051~Oph
	that agree very well with previous estimates. Furthermore,
	they found no
	discrepancy to previous distance estimates for the systems
	IP~Peg and VZ~Scl. This gives us confidence in our method,
	though naturally, it must be tested on more systems.

\end{itemize}

As the method itself cannot clearly be blamed for the discrepancy, 
we must find another solution to this discrepancy tied to
the structure of the disc itself.

\subsection{The solution: a patchy disc}
\label{patchydisc}

An attractive way to reduce the flux from the disc without
substantially changing the colours and total disc area is by allowing
for a covering factor: the disc need not be uniformly covered with
emitting material if only a fraction of the local disc surface
can produce the observed optical flux.  Since the ``chromospheric''
emission modelled here is also seen in the near infrared in other
quiescent dwarf novae (Berriman et al.\ 1985), the remaining area must
be filled with ``dark'' material which may only show up at even longer
infrared wavelengths, i.e.\ dark patches which are invisible in the
optical part of the spectrum. The covering factor we require is about
$C = (133/207)^2=41$\%.

While the presence of such a covering factor $C < 100\%$ is a
plausible solution of the distance discrepancy, it is at present not
possible to establish if and how the covering factor may change with disc
radius or azimuth. In future analyses where the distance to a CV is
well known one may consider adding a covering factor $C$ as an
additional parameter to be mapped. This may be especially
viable if we use eclipses of the line emission.

Fortunately, the physical parameters of the patchy disc located at the
(true) $d \sim 150$~pc will not be very different from those
reconstructed at $d \sim 200$~pc since the total subtended area of
the emitting regions (as opposed to the total disc area) is the
same. {\em This preserves all of the important spectral properties
except the effective temperatures, $\Teff$}. The latter are derived
from the bolometric fluxes and the total (rather than the emitting)
areas; the $\Teff$'s need to be scaled down to match the lower
distance by a factor of $(0.41)^{1/4}=$80\%.  The resulting
mass-accretion rates in the maps are thereby lowered (as can be seen
in Fig.\,\ref{htteff}) to values below or at least closer to the
critical mass-accretion rate.

By moving HT\,Cas closer, the fitted temperature of the white dwarf is
also reduced.  In Appendix~\ref{app_cover} we show a reconstruction with
a fixed covering factor $C = 41\%$ for a distance of 133pc. The resulting
white dwarf temperature is 15\,500~K.   For 150~pc and $C = 52\%$, we
get 17\,450~K, close to Wood et al.'s (1995) value of
$18\,700\pm 1\,800$.

The $\alpha$ values will also be reduced for such a patchy disc, but
only down to values of 10-100. However, the presence of additional but
optically unseen material solves the problems associated with the
still unreasonably large $\alpha$ values: if most of the viscous
transport occurs in the unseen material and the small amount of
observed ``chromospheric'' material finally radiates away this
dissipated energy, then reasonable values of the ``turbulent''
viscosity can be invoked.  The relative amount of unseen matter is
then at least equal to the inverse of the derived $\alpha$, i.e., of
the order of tens or hundreds of grams per cm$^{2}$.  As theoretical
calculations (e.g.\ Ludwig et al.\ 1994) show, the hysteresis curve in
the effective temperature - surface density diagram used to explain
dwarf nova outbursts lies just at these surface density values.

Thus, if at optical wavelengths we only see the chromosphere, we
underestimate the surface densities severely and therefore
overestimate the $\alpha$'s. A simple estimate shows that just such
surface densities of 10-100 g~cm$^{-2}$ in an underlying cool
disc lead to the desired values of $\alpha < 1$.

An analysis of single eclipses (see Fig.\,1 in HWS91) could in principle
reveal the pattern, however, the pattern would be smeared out due to
the differential rotation of the disc. Another problem one encounters
with mapping single eclipses is the flickering, easily hiding just
such small steps expected in the eclipse profile. The flickering
introduces features in the map that cannot be distinguished from real
patterns of a patchy disc.

Our analysis therefore reveals the disc as an average between the
bright and dark patches. If the dark patches are so cool that they are
basically undetectable at the optical wavelength at which the data
were taken, the spectra (colours) in the disc are determined by the bright
patches. A time average, however, is lower in intensity than that of
an active region, due to the averaging over times when the patch is
present and absent. This leads to the increase in distance that we
derive using PPEM.

\section{The source of the patchiness}
\label{sourcepatch}

The PPEM reconstructions of the intensity distribution of the disc in
HT\,Cas imply that the optical emission does not cover the entire
face of the disc.  If proven, this fact would place severe constraints
on the implied structure of the disc.
In this model, most matter is carried by an underlying optically
thick, cool accretion disc which generates the energy finally
radiated away in the chromosphere. This underlying disc must have
surface densities of the order of tens to hundreds of g\,cm$^{-2}$ 
in quiescence in order to explain the dwarf nova eruptions
(e.g.\ Ludwig et al. 1994).

The presence of patchiness would argue against models for chromospheric
emission which is simply produced either by irradiation-induced coronae
(Smak 1989) or thermally unstable viscous atmospheres (Adam et al.\ 1984),
since these processes should produce a uniform chromospheric layer over
the entire geometrically thin disc.   This argument could be weakened
if the disc were able to maintain a vertically ``corregated'' structure
like that suggested in systems with spiral waves (Steeghs, Harlaftis
\& Horne 1997), in which case either irradiation or locally viscous
and thermal processes could produce the required mottled emission.
For example, the opening angle of spiral waves in discs is supposed
to decrease with decreasing azimuthal Mach number, so quiescent
discs may have very tightly wound-up spirals which either produce
local dissipation or raise material high enough to be irradiated by
the central disk/star.

An alternative model is that the patches are similar to active regions
on the surface of the sun and could be linked to the flickering in the disc.
These regions could be due to magnetic activity in localized regions on
the surface or in the upper layers of the accretion disc.  For example,
magnetic flux created by dynamo action and/or the Balbous-Hawley
instabilities driving the viscosity would rise out of the cool midplane
regions and dissipate most of the viscously generated energy via magnetic
reconnection or similar coronal processes.  This model would also explain
why the emission line intensities are proportional to the local
orbital frequency (Horne \& Saar 1991).

The previous models generally assume that the optical emission region
is an upper layer above an optically invisible disc containing most
of the mass, even locally along a vertical cut through the disc.
A very different alternative is suggested by the observation of
``mirror'' line eclipses in the Paschen lines in the dwarf nova IP\,Peg
by Littlefair et al. (2001) at orbital phases around 0.5.  Such behavior
can only be explained by having a disc which is optically thin enough
to let the light of the secondary star shine partially through the disc
at phase 0.5.  This observation is inconsistent with a uniform, optically
thick cold disc with surface patches of chromospheric emission on top
(and bottom) and suggests (1) that the patches implied by the
PPEM reconstructions are exactly those areas of the disc which are
vertically truly optically thin and
(2) that the dense, cold material needed to
explain both the surface densities necessary to explain the dwarf
nova eruptions and the anomalous
derived viscosity parameters is laterally rather than vertically
distinct from the chromospheric patches.
The fact that the ``mirror'' eclipses are seen at large Keplerian
velocities argues against patchiness which is simply due to the
``shredding'' of the outer disc by the accretion stream (weak in
HT\,Cas) and the secondary's tidal forces.  Thus, there would have
to be some thermal or viscous instability which produces the mottled
structure throughout the disc.

\section{Lifetime and size of patches}

If the patches are associated with reconnection events (disruption of
structures) due to MHD turbulence, the timescales for the creation of
structures will be linked to the local Keplerian timescales ($t_K(r)$).
\begin{eqnarray}
t_K(r) & \equiv & 2\pi/\Omega_K(r)\nonumber\\
       & =  &~ 18 ~ \left( \frac{M_{wd}}{M_\odot} \right)^{-\frac{1}{2}}
                         \left( \frac{r}{10^9\,cm} \right)^{\frac{3}{2}} 
                        {\rm sec} \nonumber
\end{eqnarray}
For HT Cas, $t_K(r) \sim$ 10sec
near the white dwarf and $\sim$ 17min ($<P_{orb}$) at the disc edge
(0.4$\Rl$).

The timescales for reconnection events can be estimated using the
timescales for flickering -- between a few tens of seconds and several
minutes (Bruch 1992), i.e. covering the full range of Keplerian
timescales in the disc.  Flickering could either be caused by
temperature fluctuations on the Kepler timescale (Welsh, Wood, Horne
1996) or caused by a self-organized criticality (or avalanche) effects
as described by Yonehara, Mineshige \& Welsh (1997).

The maximal lifetime of a magnetic structure is determined by the differential
rotation of the disc material. For example, whenever a magnetic loop
is rooted at different radii, $r$ and $r+\Delta$r, shearing in the disc
will tend to destroy its coherence on a timescale
\begin{eqnarray}
t_{\rm co} & \approx &
        \left(
        t_K(r)^{-1} - t_K(r+\Delta r)^{-1}
        \right)^{-1} \nonumber\\
 ~ & \approx & \frac{4\pi}{3} t_K(r) \frac{r}{\Delta r} \nonumber\\
 & \approx & 13 \left( \frac{M_{wd}}{M_\odot} \right)^{-\frac{1}{2}}
                  \left( \frac{r}{10^9\,cm} \right)^{\frac{3}{2}} 
                  \left( \frac{\Delta r/r}{0.1} \right)^{-1} \nonumber
                  {\rm min}
\end{eqnarray}
$t_{\rm co}$ is of the order of a few minutes in the inner disc and
several hours (i.e.\ several hours) at the disc edge.

The size of the structures is linked to the disc thickness at
each radius. If the vertical and radial velocities in the disc are
comparable, then the radial size may be comparable to the disk
thickness.  The azimuth size of the structure will be somewhat larger
due to differential rotation shearing out the structure.

It will probably be very difficult to map the structures with
photometric light curves, because single eclipses will contain the
projection of many structures. These cannot be resolved
2-dimensionally with eclipse durations $\gg$ the flickering timescale,
because the flickering prevents us from seeing the signatures of the
structures in the eclipse profile. Furthermore, the patches are moving
on a Kelper time scale generally (much) smaller than the eclipse
duration which leads to smeared out patches.  Averaging over several
cycles will average over times when the structures are at different
places (see also Section 7.2). Thus, we can only expect to see the
structures in fast and high resolution spectroscopy. The patches will
lead to fine-scale wrinkles that will vary with the local Kepler
velocity along the emission line profiles (Hoffmann, Hessman \&
Reinsch 2002).

\section{Summary}

We have applied the {Physical Parameter Eclipse Mapping} (PPEM) method
to archival UBVR photometry of the eclipses in the quiescent dwarf nova
HT~Cas in order to map the optically thin matter responsible for the
optical continuum and emission line spectrum.

Using a spectroscopically determined distance, we were unable to
obtain reasonable maps of the temperature and surface density in the
disc: the reconstructions showed characteristic artefacts showing that
the surface brightness of the disc was too high.  By insisting that
the disc be maximally structureless, we were able to find reasonable
solutions at a distance of 207pc.  The resulting accretion disc has a
radius of 0.3-0.4$\Rl = 9-13 \times 10^9$~cm, kinetic temperatures
between $4000$~K and $10^4$~K, surface densities between 0.013 and
0.04\,g\,cm$^{-2}$, and $\tau \lsim 1$ everywhere except near the hot
central white dwarf (22\,600~K).  The effective temperature is flat in
the inner part of the disc and may follow a steady state solution
towards the disc edge.  We derive apparent values of the viscosity
parameter $\alpha$ from the bolometric fluxes, temperatures, and
surface densities which are orders of magnitude higher than they
should be.

Acknowledging that the distance to HT~Cas is unlikely to be as high as
suggested by our PPEM entropy analysis, we explain why
the problem is unlikely to be due to a poor treatment of the white
dwarf or the wrong choice of a spectral model.
We are able to rectify both the poor reconstructions of the models
using $d \sim 150$~pc and the problems with those using
$d \sim 200$~pc
by invoking emission regions which are not distributed uniformly
across the face of the disc but are patches with a
covering factor of only 41\%.  Since the angular extents of the
emitting regions in the distant non-patchy and close by patchy discs can
be identical, the reconstructed physical parameters of the $d \sim 200$~pc
solutions are still valid at $d \sim 150$~pc
with the exception of those quantities which depend upon the
bolometric fluxes: the mass-accretion rates (or effective
temperatures) need to be scaled
down by 80\%. This reduces the local mass-accretion rates to just at
or below the critical rates for maintaining the dwarf nova eruptions
for all but the innermost disc.

The empirical values of $\alpha$ (cf.~\ref{viscosity}) -- ranging
from about 10 to 1\,000 -- for both $d \sim 150$~pc and $d \sim 200$~pc
are still orders of magnitude larger than is physically plausible.
Fortunately, the patchy chromosphere on the disc surface requires
a substantial mass of very cool material which powers the
viscous energy finally radiated away in the chromosphere and which
have surface densities of the order of tens or hundreds of
g\,cm$^{-2}$ -- just those needed by the disc instability models.

The patchiness excludes models for disc coronae due to local thermal
instabilities or to irradiation.
If we interpret the patchy chromospheric emission regions as
magnetically active regions, the fact that the dissipation is proportional
to the angular velocities could be a sign that what we are seeing
magnetic flux created by magnetohydrodynamical instabilities and/or
disc dynamos responsible for the anomalous viscosity in discs
dissipated in magnetically active regions. 
However, the presence of ``mirror eclipses'' in another dwarf nova
implies that the warm chromospheric patches may be situated laterally
adjacent to rather than simply above the cold, more massive regions.

HT~Cas may really deserve its name as ``Rosetta Stone''
of dwarf novae, since it could be the key to understanding the
anomalous viscosity in accretion discs as due to hydromagnetic
turbulence. High time and spectral resolution spectroscopy using a
10-m class telescope and/or multi-colour observations in a yet to be
detected low state could provide the final proof for our hypothesis.

The Physical Parameter Eclipse Mapping proves to be a powerful tool to
investigate the physics of accretion discs -- even when it at first
appears to produce a wrong result!

\section*{acknowledgments}

We thank Brian Warner for fruitful discussions and Klaus Beuermann for
intensive discussions on the determinations of distances to
CVs. Furthermore, we thank the anonymous referee for valuable comments
that led to an improvement of the paper. SV thanks the European
Comission and the South African NRF for funding through fellowships.

\begin{appendix}
\section{Test for Distance estimate}
\label{app_dist}

In order to test the reliability of the PPEM distance estimate, we performed a test
in which we calculated a light curve from an arteficial disc and
attempted to reconstruct it using various different trial distances.

\begin{figure}
\hspace*{0.5cm}
\psfig{file=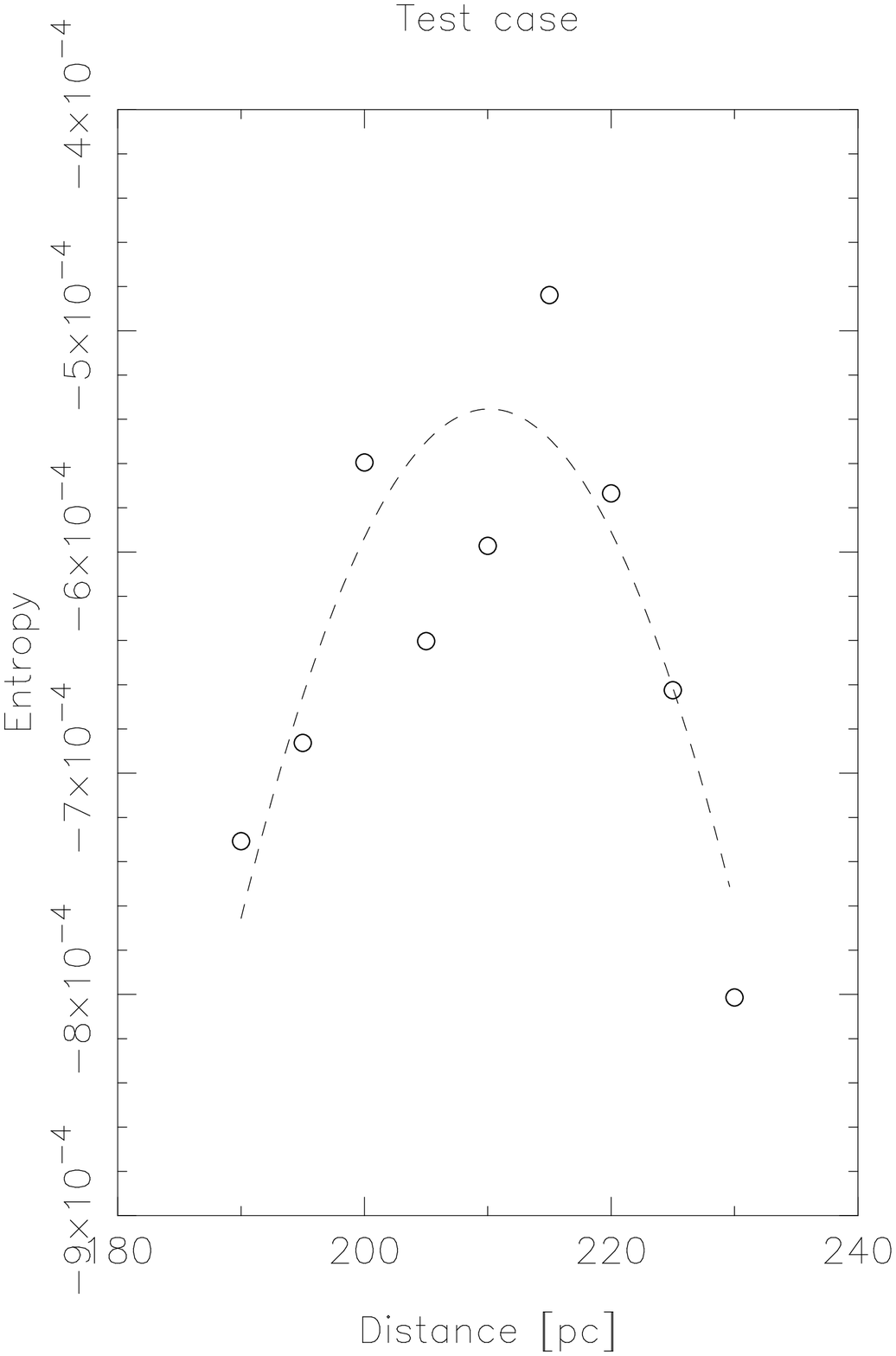,width=8cm}
\caption{\small The entropy - distance relation for the test case.
For each trial distance the data were fitted with
$\chi^2/N = 1$. The dashed line is a parabolic fit to the data,
peaking at 210~pc.
\label{disttest}}
\end{figure}

Our test disc consists of an axisymmetric temperature and surface
density distribution with a bright, gaussian spot at a distance of
$0.3\Rl$, an azimuth of 20$^\circ$ and a sigma of 0.02$\Rl$. We added
this non-axisymmetric feature in order to test if this will influence
the distance estimate.

\begin{figure*}
\hspace*{0.1cm}
\psfig{file=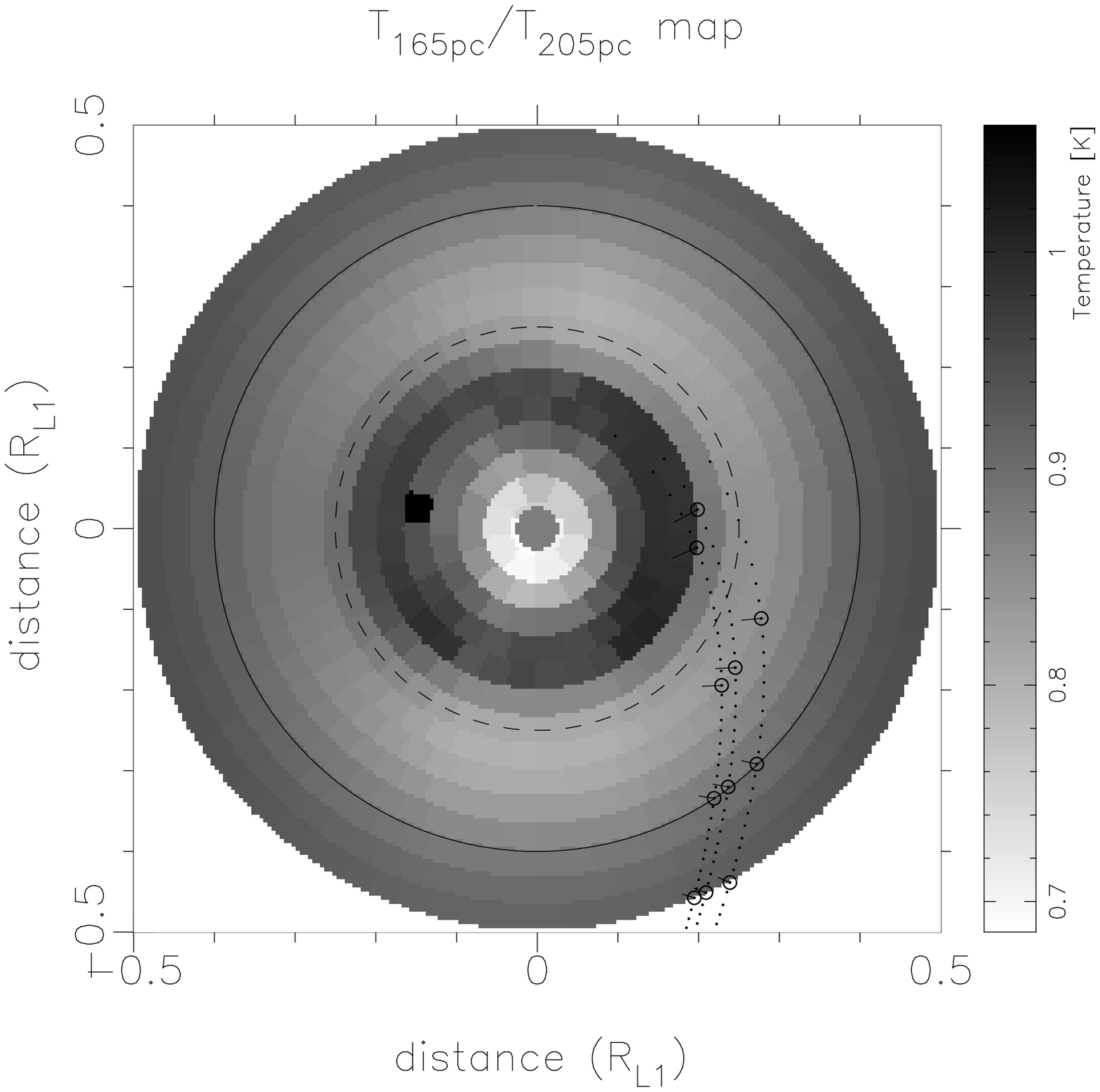,width=8cm}
\hspace*{0.1cm}
\psfig{file=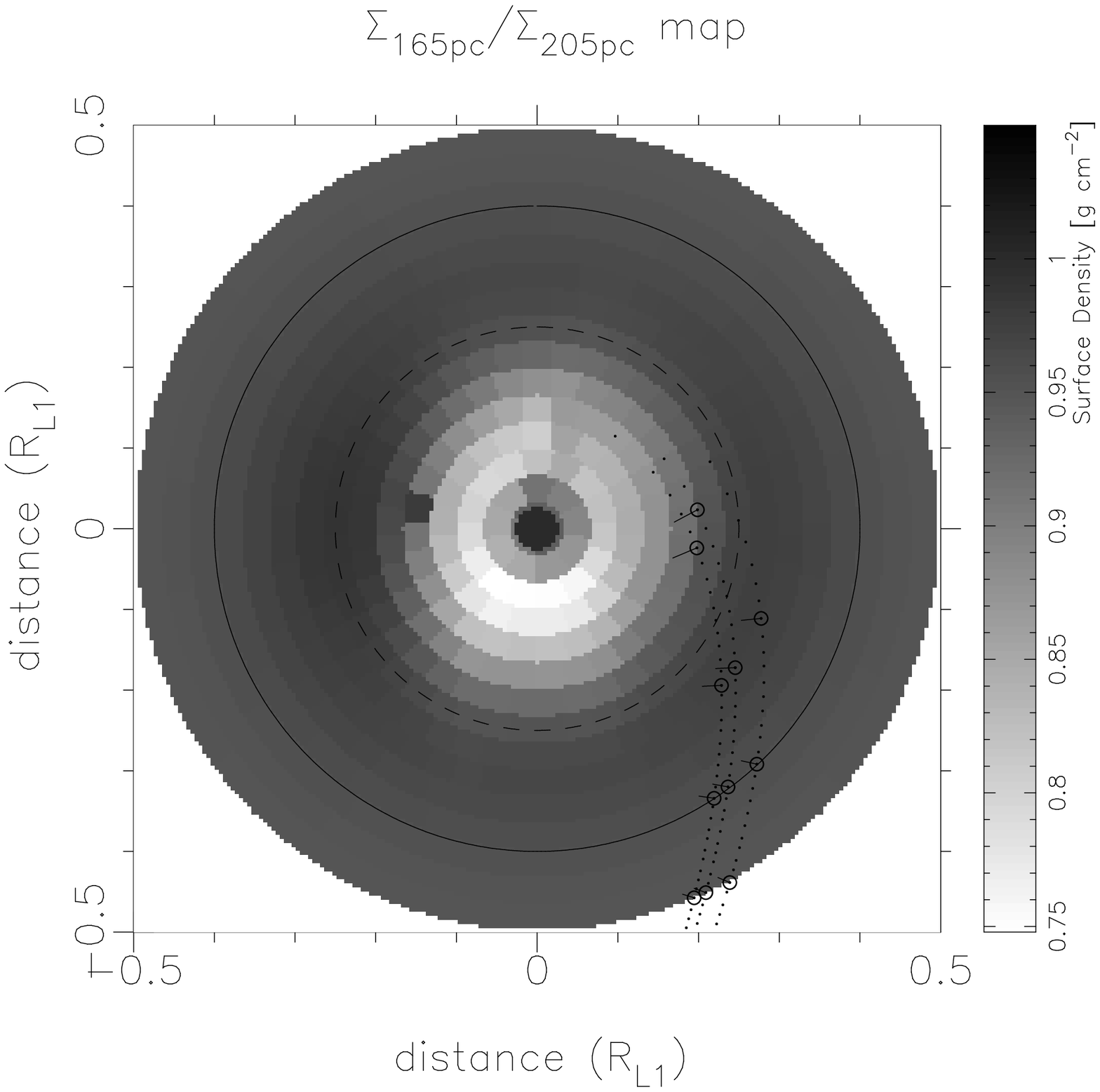,width=8cm}
\caption{\small The ratio map in temperatures ({\em left}) and surface
densities ({\em right}) for trial distances 165~pc (Wood et al.\ 1995)
and 205~pc (PPEM, see Section~\ref{distestimate}).  The Roche-lobe of
the primary component lies just about outside the plotted region.  The
dashed line at 0.25$\Rl$ and the solid line at 0.4$\Rl$ are drawn for
comparison with Fig.\,\ref{htmaps}. The curved, dotted lines in the
grey-scale plot are theoretical accretion stream paths for mass ratios
0.15 $\pm$ 50\%. The secondary is at the bottom.
\label{htr165}}
\end{figure*}

The artifical light curves were calculated using the same phase
resolution as for the HT Cas data displayed in Fig.\,\ref{htf}, i.e.
0.003 outside the eclipse (phases $\pm$0.065) and 0.0015 during
eclipse yielding 112 data points between phases $\pm0.1$. We added
noise with a signal-to-noise level of 100.
The correct distance was assumed to be 205~pc.

Fig.\,\ref{disttest} shows the resulting entropy - distance
relation. A parabolic fit to the data points peaks at 210~pc, i.e. at
a distance 5~pc larger than the true distance.

In several applicaton we noticied that when fitting light curves, the
distance decreases with decreasing $\chi^2$ (see e.g.\
Fig.~\ref{dist}). This effect is usually only small, about 5~pc for a
drop in $\chi^2$ of 1. In applications to real data, one usually
encounters a lower limit of $\chi^2$ below which no good fit to the data can
be found (or the reconstructions start showing artefacts). This limit
is determined by flickering in the data or by the adequacy of the spectral model
used (for a discussion on the appropriateness of our
model see Section~{resolution}).  We expect therefore, that the true
distance will be slightly smaller than the derived distance.

\section{Reconstruction with smaller distance}
\label{app_short}

To illustrate the difference between the maps for a trial distance of
165~pc as determined by Wood et al.\ (1995) and our PPEM distance
estimate we plot the ratio of the parameter maps in
Fig.\,\ref{htr165}. To allow comparison, we used the reconstructions
for a $\chi^2$ of 2.5. An asymmetry can clearly be seen. Since the map
for 205~pc is much smoother than the one for 165~pc, the asymmetry is
mainly due to the latter map.

The temperature ratio map shows maxima in two areas at azimuths
$\pm90^\circ$, while the surface density ratio map shows a minimum at
azimuth $0^\circ$. This shows that the algorithm tries to maintain the
eclipse width and at the same time to reduce the emitting area, leading
to a shortening of the disc along the binary axis and thus to an
elongated emission structure perpendicular to the binary axis. Such an
emission structure is usual for a reconstruction for which the trial
distance is too small.

Additionally, the reconstruction shows a single hot pixel at
approximately azimuth $-95^\circ$ and radius 0.15$\Rl$. Such hot
pixels are usually artefacts, since it is physically unlikely to
maintain such a sharp gradient in temperature and in such a small
area. It is especially unlikely since the light curves are
averaged. Such a hot pixel is usually an
additional indication for a false trial distance.

\section{Reconstruction with larger distance}
\label{app_long}

\begin{figure*}
\hspace*{0.1cm}
\psfig{file=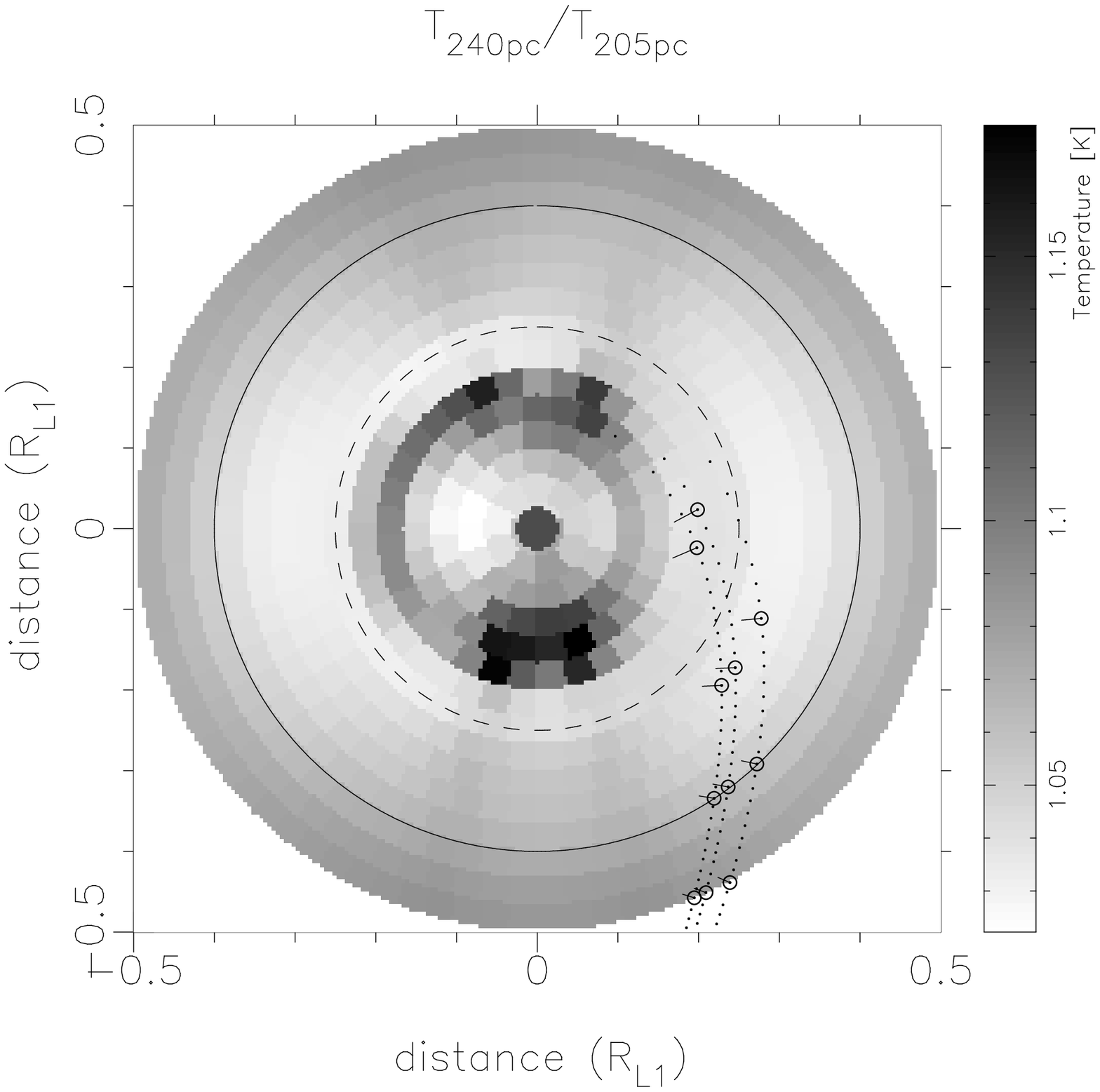,width=8cm}
\hspace*{0.1cm}
\psfig{file=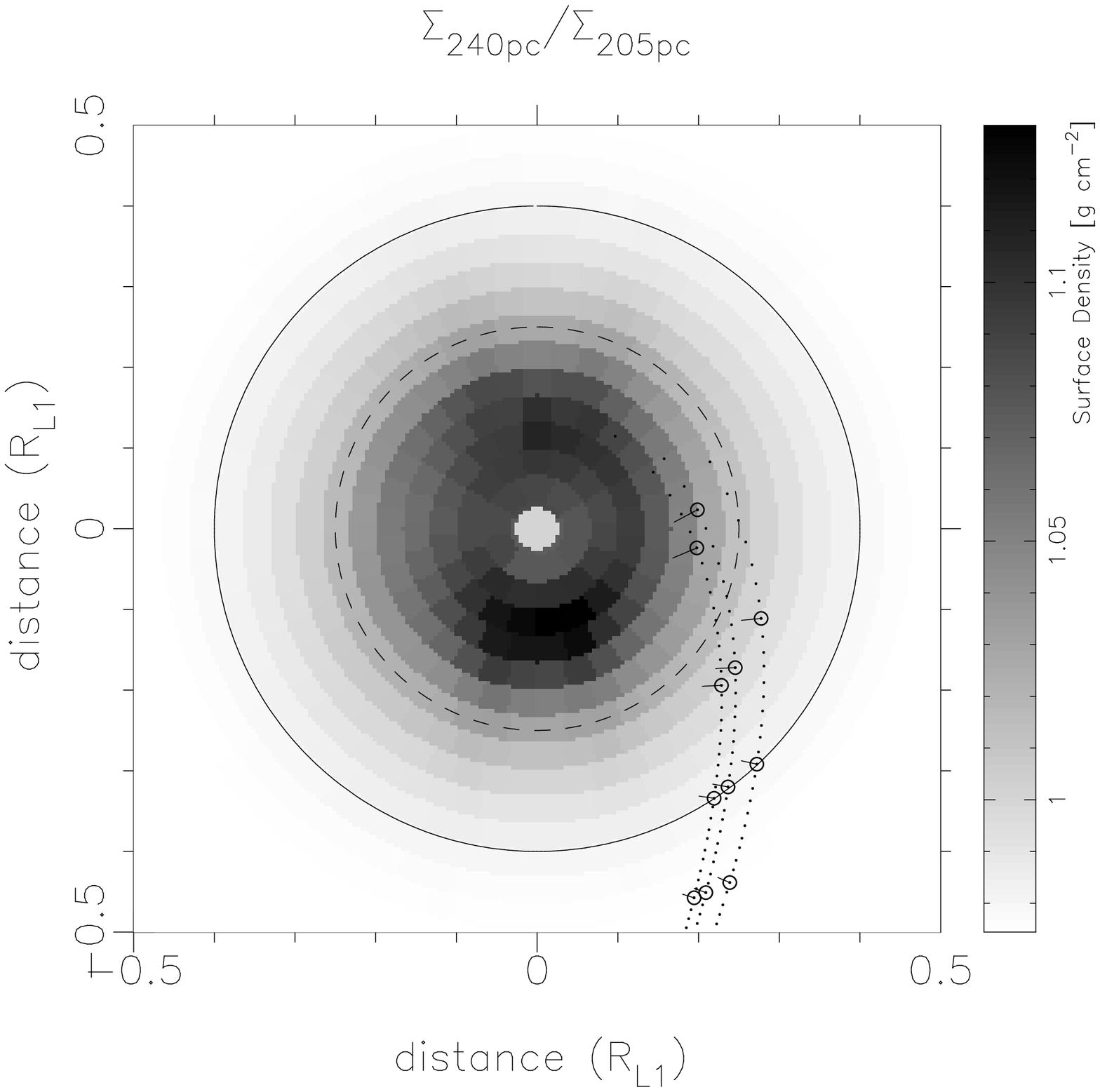,width=8cm}
\caption{\small The ratio map in temperatures ({\em left}) and surface
densities ({\em right}) for trial distances 240~pc (Wood et al.\ 1995)
and 205~pc (PPEM, see Section~\ref{distestimate}).  The Roche-lobe of
the primary component lies just outside the plotted region.  The
dashed line at 0.25$\Rl$ and the solid line at 0.4$\Rl$ are drawn for
comparison with Fig.\,\ref{htmaps}. The curved, dotted lines in the
grey-scale plot are theoretical accretion stream paths for mass ratios
0.15 $\pm$ 50\%. The secondary is at the bottom.
\label{htr245}}
\end{figure*}

For completeness we show the difference between the reconstructions
for trial distances 240 and 205pc as a ratio of the parameter maps in
Fig.\,\ref{htr245}. Again we used a $\chi^2$ of 2.5 for comparison.
These maps show also an asymmetry, however, a distinctly different
pattern. Since the map for 205pc is very smooth for a $\chi^2$ of 2.5,
the asymmetry is mainly due to the map for 240pc.

The maps show maxima along an axis parallel to the binary axis. This
can be understood if one considers the following: the eclipse width is
defined by the eclipse profile and determines the extension of the
accretion disc perpendicular to the binary axis. However, if the
distance is too large, the disc needs to cover more area than
available if the disc were axisymmetric. Therefore, the emission
region in the disc extends along the binary axis without compromising
the eclipse width.  This pattern is thus typical for a map that was
derived with too large a trial distance.

The reason for the enhanced structure in the disc is the same as for
the curious hot pixel in the map reconstructed with the small
distance.

\section{Reconstruction with partially visible white dwarf}
\label{app_wd4}

\begin{figure*}
\hspace*{0.5cm}
\psfig{file=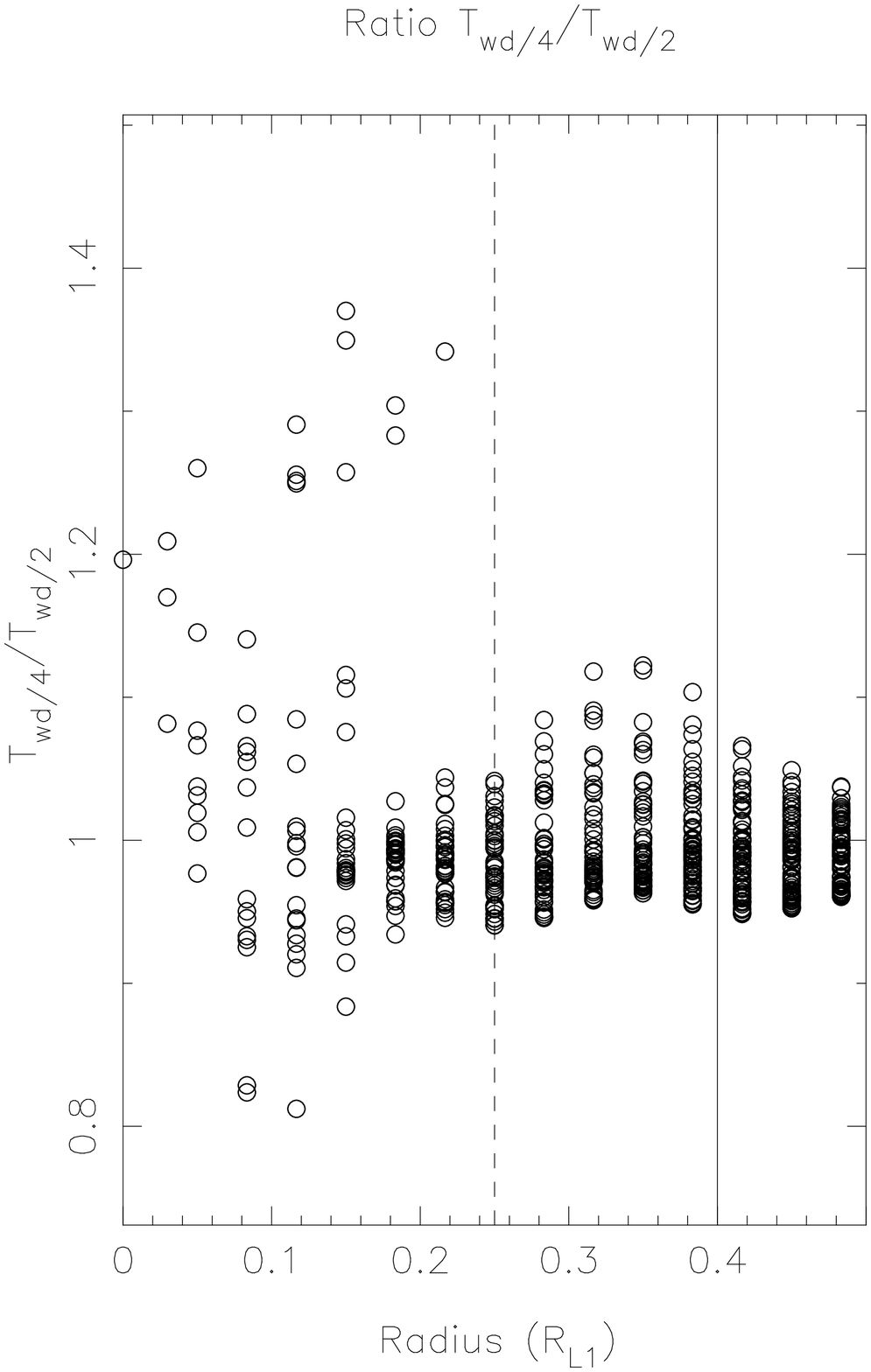,width=7cm}
\hspace*{1cm}
\psfig{file=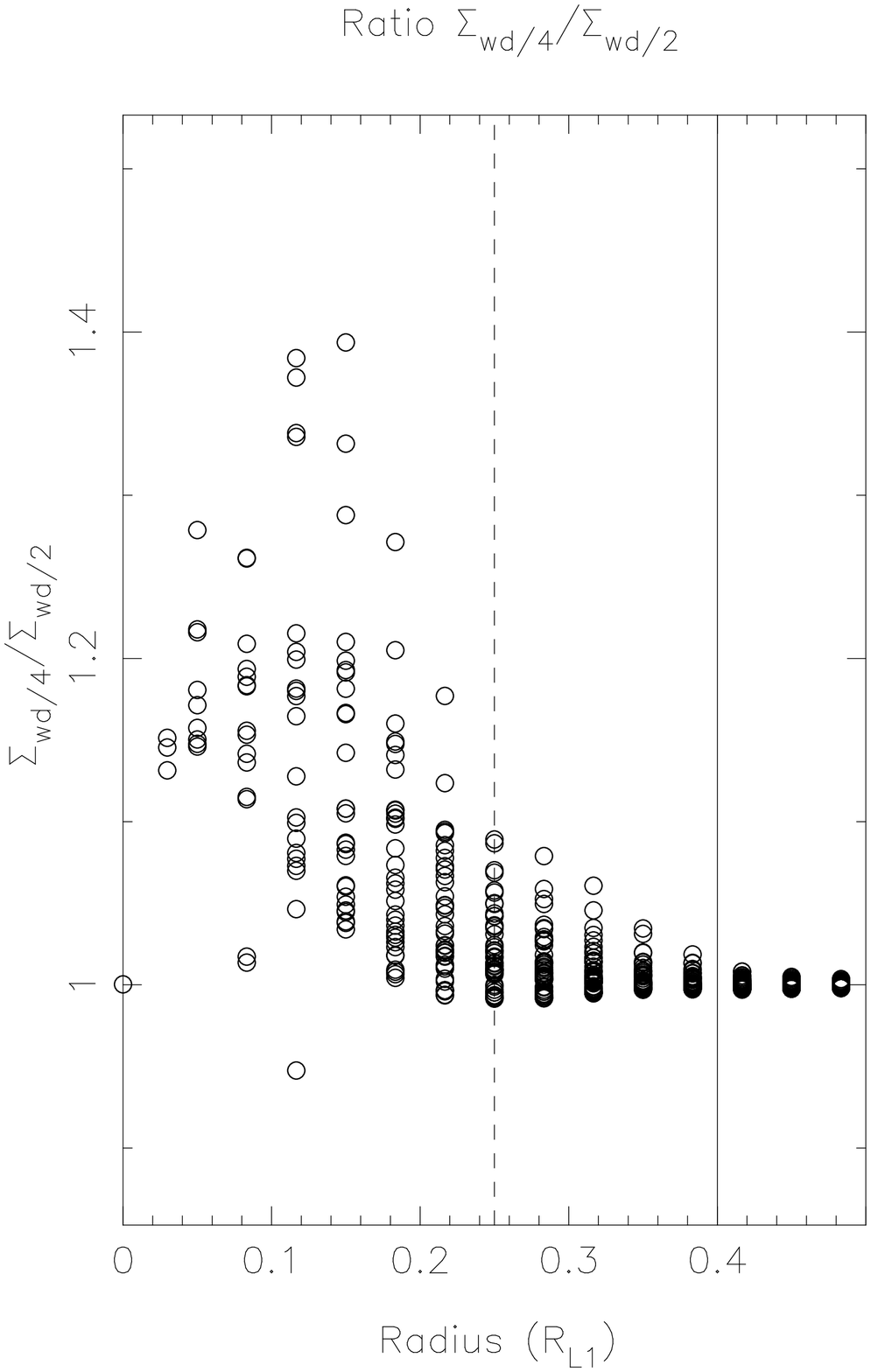,width=7cm}
\caption{\small The ratio of the temperature and surface density
reconstructions using a partially visible white dwarf ($T_{wd/4}$,
$\Sigma_{wd/4}$) and a fully visible white dwarf ($T_{wd/2}$,
$\Sigma_{wd/2}$). The dashed line at 0.25$\Rl$ and the solid line at
0.4$\Rl$ are drawn for comparison with Fig.\,\ref{htp}.
\label{htr}}
\end{figure*}

Since the inner part of the disc appears to be optically thick, the
``lower'' hemisphere of the white dwarf should be (at least partially)
occulted. Therefore, we reconstructed the disc and white dwarf
temperature with presetting the occultation of the central object.

These reconstructions show significantly more scatter than the ones
shown in Fig.\,\ref{htp} for the fully visible white dwarf. This
scatter is expressed by a much lower entropy of these maps compared to
the ones using a fully viaible white dwarf: the value of the entropy
is $-1.1\times10^{-3}$, a factor of 3.5 below the one for the fully
visible white dwarf and outside the plotted region in
Fig.\,\ref{dist}. Therefore, the PPEM reconstruction favors the fully
visible white dwarf, like Horne \& Wood (1990).

Fig.\,\ref{htr} shows the ratio of the two reconstructions. As
expected, the white dwarf temperature in the case of the white dwarf
with a smaller visible area is higher ($T_{wd,1/4} = 26\,700$K). In
addition to showing more scatter, the values of the disc temperatures
and surface densities are generally higher. This is due to a combined
attempt to reproduce sufficient total flux and the correct colours,
since a slightly hotter white dwarf shows a slightly bluer spectrum.

\section{Reconstruction with various default maps}
\label{app_default}

\begin{figure}
\hspace*{0.5cm}
\psfig{file=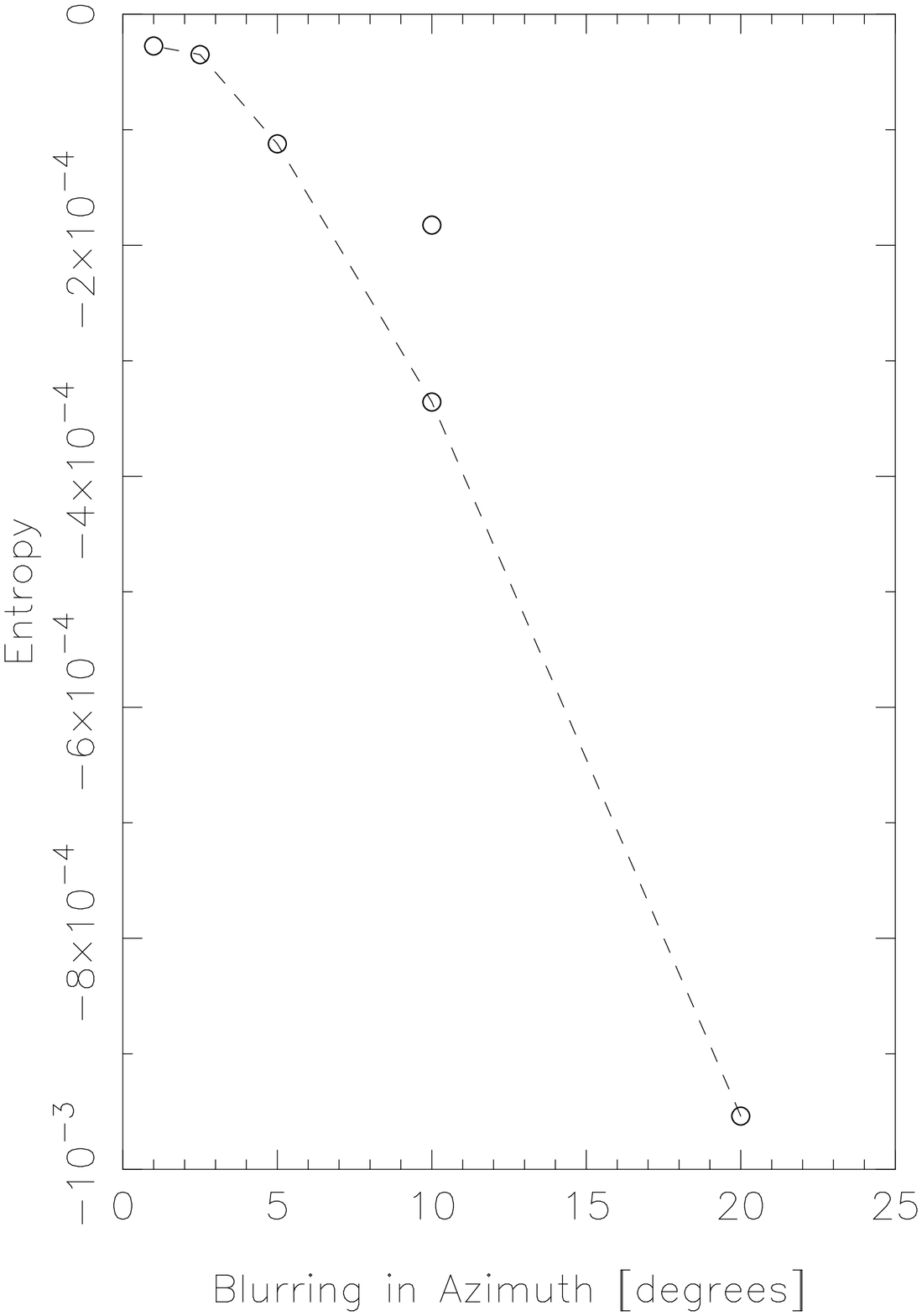,width=7cm}
\caption{\small The entropy for reconstructions with various blurring
parameters. The plot gives only the degree of blurring in azimuth,
Tab.~\ref{tab_blurr} also gives the amount of blurring in radius. The
circles connected by the dashed lines are for solid arc smearing, the
single point corresponds to solid angle smearing. The entropy for {\em
huge} $360^\circ$ blurring is $-7.7\times10^{-4}$ (it is omitted from
the plot for clarity).
\label{testblurr}}
\end{figure}

During the iteration of an eclipse mapping method like PPEM, the
intermediate map is compared to a {\em default} map which in turn is
adjusted at certain, predefined intervalls. The general use of a
default map is explained in more detail in VHH99. In short: a default
map is a smeered out version of an intermediate map calculated using
certain blurring parameters in radial and azimuthal
direction.

The default map therefore introduces a certain degree of symmetry or
rather smoothness. The degree of smoothness depends on the value of
the blurring parameter. It is important to note that at the usual
blurring parameters (see below) asymmetries are not completely or even
nearly completely suppressed.  It only leads to a certain degree of smoothness in
the disc. This is desireable in order to avoid artefacts to dominate
the map and to choose one out of the many possible solutions.

In order to estimate the influence of the default map on the PPEM
solution we used a set of default smearing parameters as given in
Tab.~\ref{tab_blurr}. Usually, we use a solid arc smearing, i.e. we
use a constant arc length. This leads to enhanced blurring in the
central regions. We use this type of blurring, because we expect more
large scale structures in the outer regions (e.g.\ non-circular disc,
bright spot). In one case we used solid angle blurring, leading to
enhanced blurring in the outer regions of the disc.

The reconstructed temperature ($T$) and surface density ($\Sigma$)
maps all show similar absolut values for $T$ and $\Sigma$ only with a
larger scatter in the maps corresponding to the larger blurring
parameter values. As an indicator for the random scatter,
Fig.\,\ref{testblurr} shows the entropy for the solutions using the
blurring parameters as given in Tab.~\ref{tab_blurr}.

\begin{figure*}
\hspace*{0.5cm}
\psfig{file=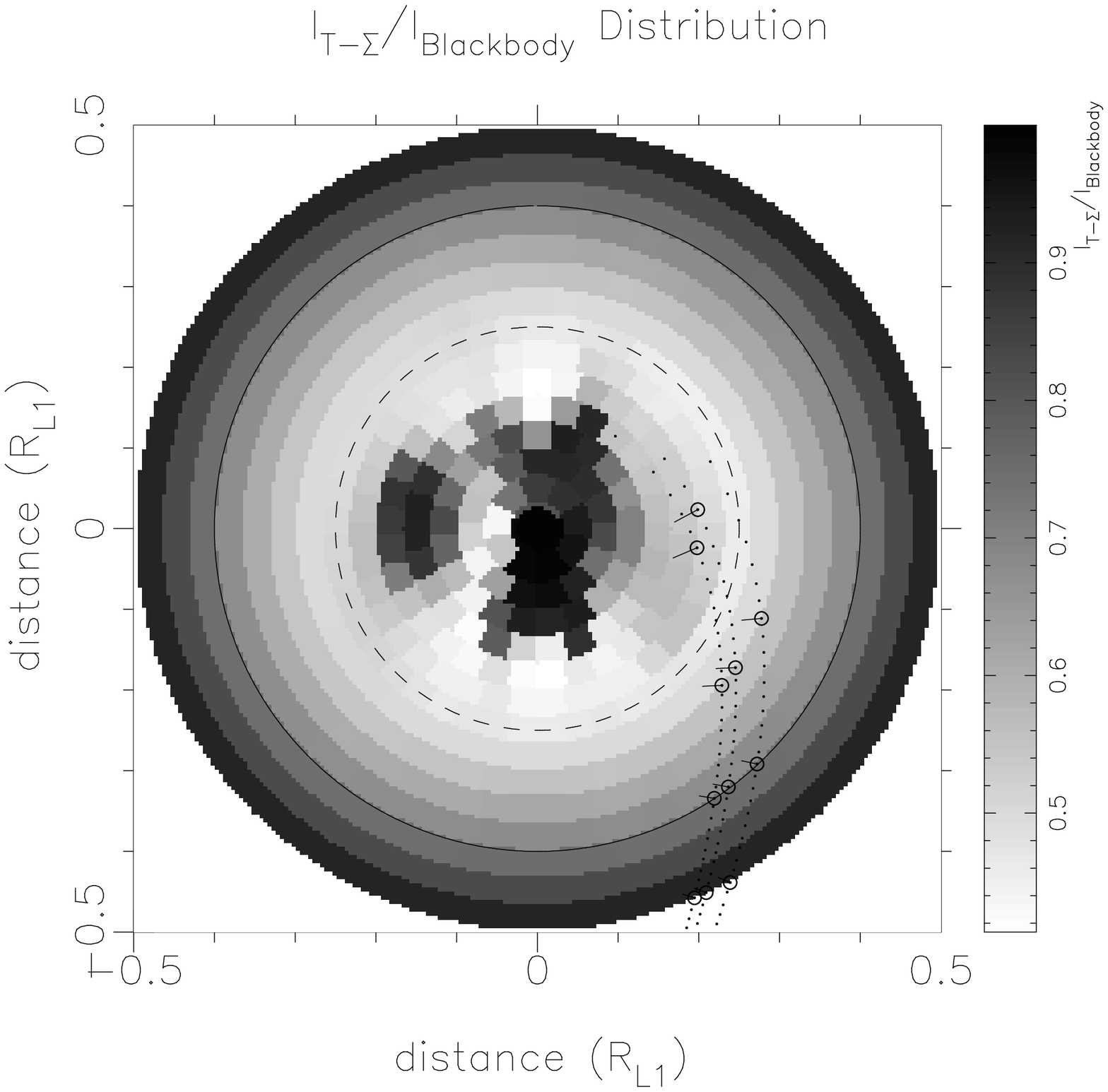,width=8cm}
\hspace*{0.5cm}
\psfig{file=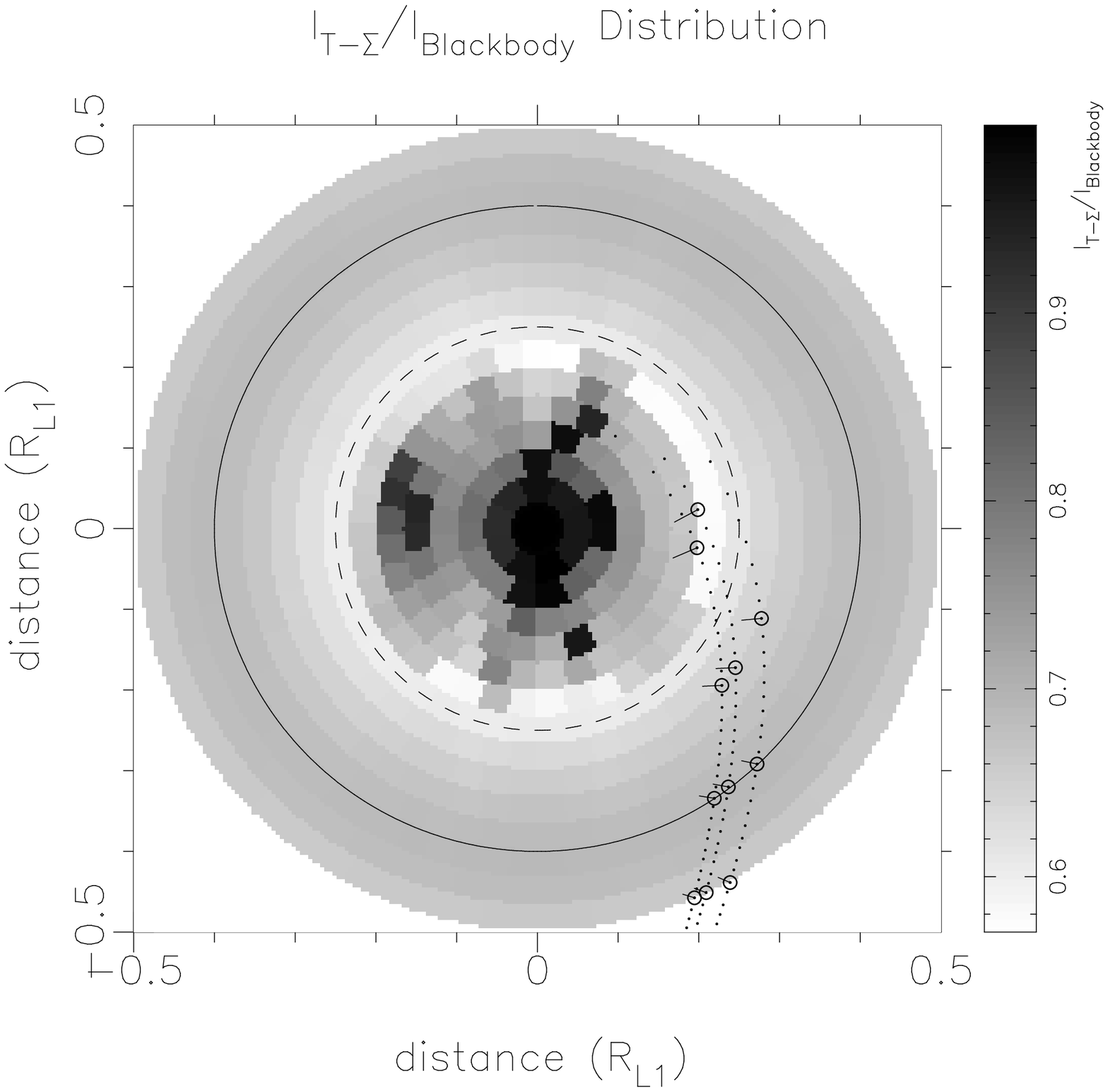,width=8cm}
\caption{\small The intensity ratio maps for {\em tiny} and {\em huge}
blurring parameters.  The Roche-lobe of the primary component lies
just about outside the plotted region.  The dashed line at 0.25$\Rl$
and the solid line at 0.4$\Rl$ are drawn for comparison with
Fig.\,\ref{htmaps}.  The curved, dotted lines in the grey-scale plot
are theoretical accretion stream paths for mass ratios 0.15 $\pm$
50\%. The secondary is at the bottom.
\label{htdef}}
\end{figure*}

As surprizing as it seems to find a higher entropy for smaller
blurring parameters, we have a simple explanation for this. Although
the maps for smaller blurring parameters seem to have more scatter
(Fig.\,\ref{htdef}), this is not random. The image in the map rather
looks more focussed, tending to separate the pixels in the map into either hot or
cool ones. Considering only the cool pixels, the disc is then very
smooth. This way, artefacts are also enhanced and can easily
be misinterpreted. We therefore prefer using intermediate blurring
parameters and have used the ones called {\em medium} for our
reconstructions.

\begin{table}
\caption{The blurring parameters. All but one default maps are calculated with
solid arc blurring. For the only exception where we applied solid angle
blurring, we used the medium blurring parameters.
\label{tab_blurr}}
\vspace{1ex}
\hspace{0.5cm}
\begin{tabular}{ccc}
description & Angle ($^\circ$) & Radius ($\Rl$)\\ \hline
tiny & 1 & 0.04\\
very little & 2.5 & 0.1\\
little & 5 & 0.2\\
medium & 10 & 0.4\\
much & 20 & 0.8\\
huge & 360 & 0.4\\
\hline
\end{tabular}
\end{table}

\section{Reconstruction with cover factor $C<1$}
\label{app_cover}

\begin{figure}
\hspace*{0.5cm}
\psfig{file=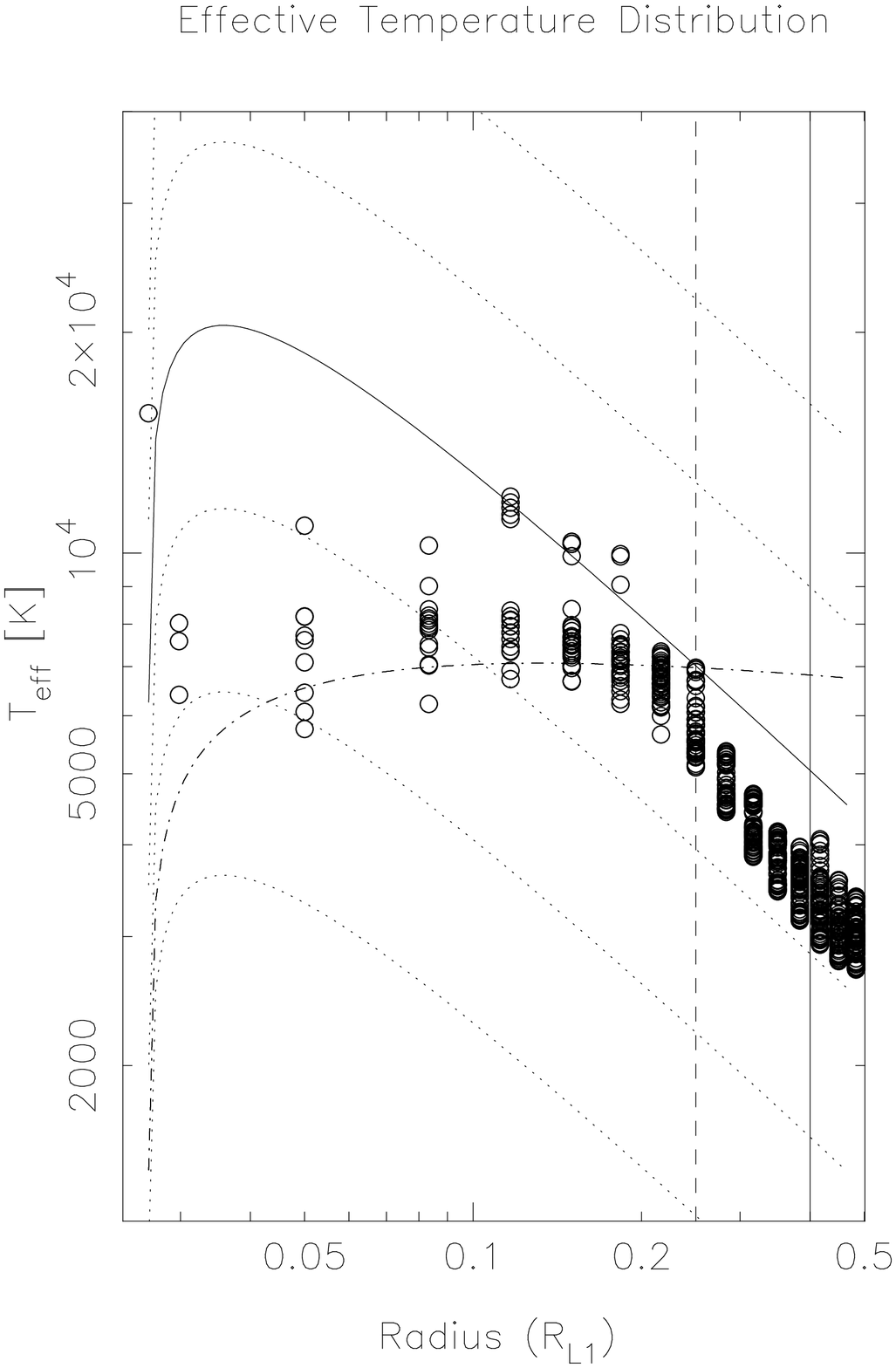,width=8cm}
\caption{\small The effective temperature reconstructed for a distance
of 133~pc and a covering factor $C=0.41$. The underlying dotted lines
are theoretical steady state temperature distributions for $\log \Md =
13$ to 18, the one for $\Md = 10^{16}$gs$^{-1}$ is drawn solid for
reference. The dashed-dotted line indicates the critical mass
accretion rates \protect{$\dot{M}_{\rm A}$} according to Ludwig et
al.\ (1994).  The dashed line at 0.25$\Rl$ and the solid line at
0.4$\Rl$ are drawn for comparison with Fig.\,\ref{htteff}.
\label{fig_cf}}
\end{figure}

\begin{figure}
\hspace*{0.5cm}
\psfig{file=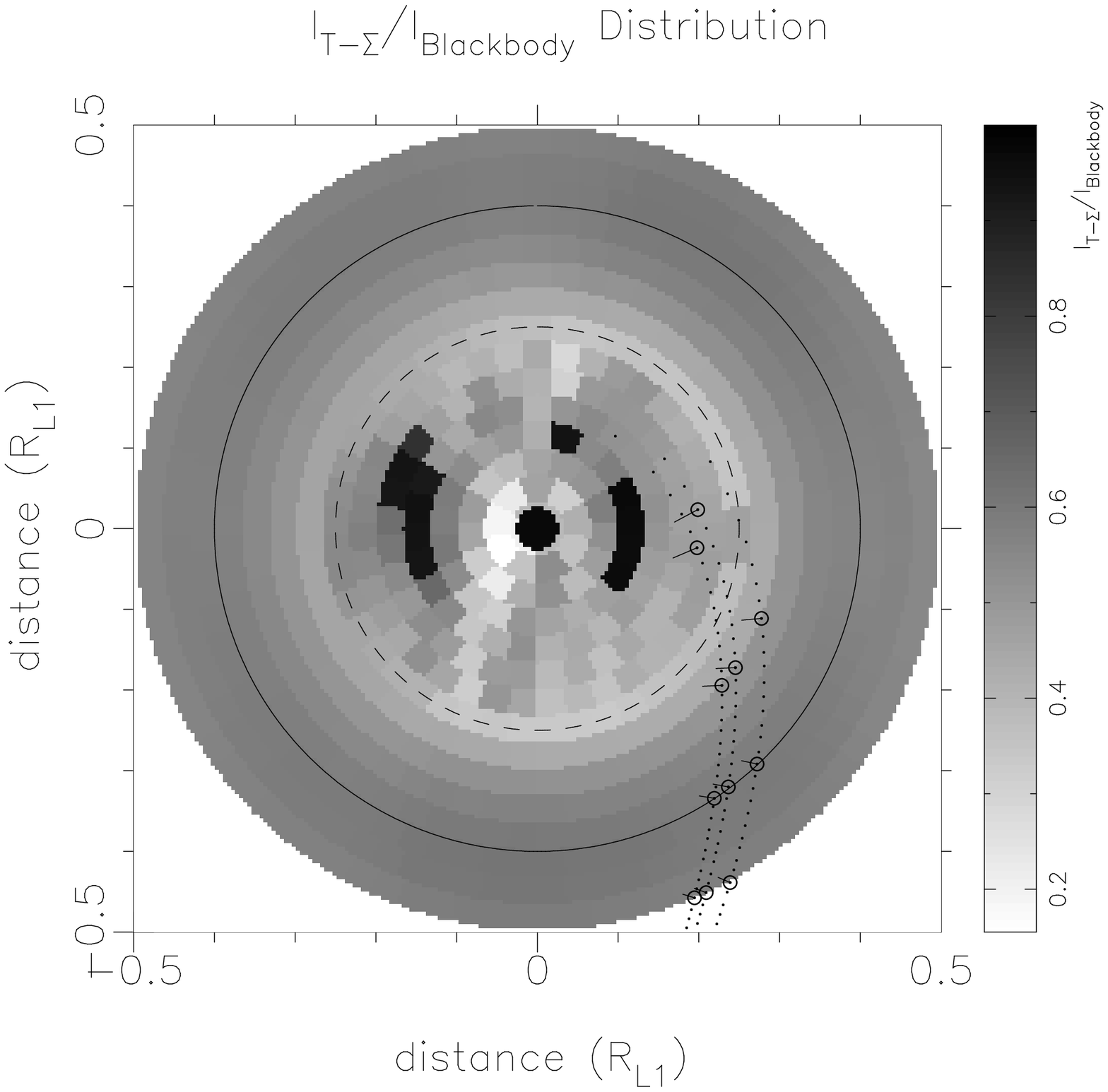,width=8cm}
\caption{\small The intensity ratio maps for a distance of 133~pc and
a covering factor $C=0.41$.  The Roche-lobe of the primary component
lies just about outside the plotted region. The dashed line at
0.25$\Rl$ and the solid line at 0.4$\Rl$ are drawn for comparison with
the plot in Fig.\,\ref{htmaps} (right). The curved, dotted lines in the
grey-scale plot are theoretical accretion stream paths for mass ratios
0.15 $\pm$ 50\%. The secondary is at the bottom.
\label{map_cf}}
\end{figure}

For completeness, we reconstructed the accretion disc for a distance
of 133~pc implementing a covering factor $C=41\%$ for the disc
emission. The resulting effective temperature ($\Teff$) map is shown in
Fig.\,\ref{fig_cf}. The $\Teff$'s range mostly below or around the
critical limit given by Ludwig et al.\ (1994), similar to the estimate
for $d=133$~pc in Fig.\,\ref{htteff}. Fig.\,\ref{map_cf}
reveals the structures in the disc. They are distinctly different to
those in maps with distances $\neq 205$pc or constructed with small
blurring parameters. We do not know whether these structures are real
or new artefacts introduced into the maps. However, it is likely that the
covering factor is not constant as assumed in this calculation and
that a reconstruction with $C$ as a parameter map will result in
more realistic solutions.

The white dwarf temperature is reconstructed to $T_{wd} = 17450K$,
compatible with Wood et al.'s (1995) estimate of $18\,700\pm1\,800$~K,
especially taking into account that they used a slightly larger distance.
\end{appendix}
\end{document}